\documentclass[final,1p,times]{elsarticle}

\usepackage{amssymb}
\usepackage{amsmath}
\usepackage{amsthm}

\begin{document}

\begin{frontmatter}



\title{Geodesic deviation equation in Brans-Dicke theory in arbitrary dimensions}


\author[1,2]{S. M. M. Rasouli}


\author[3,4]{F. Shojai}

\address[1]{Departamento de F\'{i}sica, Centro de Matem\'{a}tica e Aplica\c{c}\~{o}es (CMA - UBI), Universidade da Beira Interior, Rua Marqu\^{e}s d'Avila e Bolama, 6200-001 Covilh\~{a}, Portugal. }

\address[2]{Department of Physics, Qazvin Branch, Islamic Azad University, Qazvin, Iran.}

\address[3]{Department of Physics, University of Tehran, P.O. Box 14395-547,
Tehran, Iran}
\address[4]{Foundations of Physics Group, School of Physics, Institute for Research in Fundamental Sciences (IPM),
P.O. Box 19395-5531, Tehran, Iran}

\begin{abstract}
In this paper, we study the geodesic deviation equation (GDE) within the context of the Brans-Dicke (BD)
theory in $D$ dimensions. Then, we restrict our attention to the GDE for the fundamental observers
and null vector field past directed. Concerning the latter, in order to apply the
 retrieved GDE for the cosmological models as well as examining exact solutions, we study
 some cosmological models in general relativity (GR) and in the context of the BD theory.
 For the BD framework, we provide two appropriate settings for dynamical system, which
can be considered as the most extended case
scenarios with respect to those studied previously.
 Then, we investigate a few well-known BD cosmological models, for which we
 present the modified and corrected exact solutions.
Moreover, we show that the retrieved formalism of the
GDE can also be properly applicable for the modified BD theory (MBDT),
whose matter and potential emerge from the geometry.
For all of our herein cosmological models, we investigate
the energy conditions and depict the behavior of
the deviation vector and observer area distance.
We demonstrate that why the MBDT can be considered as a
 more appropriate candidate to describe the
universe in accordance with the observational data as well as from theoretical viewpoint.
\end{abstract}


\end{frontmatter}


\section{Introduction}

\label{int} \indent
In order to analyze the structure of spacetime geometries
for all the timelike, null and spacelike cases, the GDE could
provide an appropriate procedure. In this respect, it
has been believed that it should be considered as one of the most significant
equation in relativity, see for instance, \cite{S34,P56,P57,S87,EE97} and related papers.
Moreover, the GDE corresponding to every gravitational theory provides an
important method in order to examine the exact solutions
of the field equations.
Concretely, by applying the GDE for timelike, null and spacelike
geodesic congruences, we can retrieve Raychaudhuri equation, Mattig
relation as well as the relative acceleration of two neighbor
geodesics.

The GDE has also been studied in the alternative theories to GR: in
the context of $f(R)$ gravity, it has been studied in \cite{SS08,GCT11}.
The GDE with the Friedmann-Lema\^{\i}tre-Robertson-Walker~(FLRW)
background has been studied in the Brans-Dicke-Rastall framework \cite{SHJ16}.
In \cite{RBJF09}, by considering the induced GDE in the brane world
 models, it has been shown that proper time of test particle can be related to the integral
multiples of a fundamental Compton-type unit of length.
In \cite{DMA15}, the GDE has been studied in the context of the $f(T)$ gravity.
The GDE in the context of a generalized
S\'{a}ez-Ballester theory \cite{SB85-original,RPSM20} in arbitrary dimensions
is investigating \cite{RSM21}.

A few challenging problems in gravity/cosmology have persuaded
researchers to establish alternative theories to GR
and to generalize the standard cosmology.
Among these theories, the BD theory is one the well-known and the most studied theories.
Let us focus on the accelerated expansion of the universe in the
context of BD theory or its modifications (for a recent very
brief review, see Section I of \cite{MC19} and references therein).

In the context of the BD theory, the BD scalar
field, for a few particular cases, can play the role of the K-essence or
Q-matter and we therefore obtain accelerating scale factor at late times.
However, such models, in turn, have their own
 problems: (i) In contrary to the main idea of the standard BD theory \cite{BD61}, either a scalar
 potential\footnote{In the context of the standard
BD theory \cite{BD61}, there is no scalar potential.
Nevertheless, for later convenience, our herein generalized framework will be called the BD theory.}
must be added by hand \cite{SS01,SS03} or the ordinary matter interact
 with the scalar field. (ii) In the BD cosmological model
 investigated in \cite{BP01}, to get an accelerating universe, the BD
 coupling parameter $\omega$ must be restricted in a range which is not
 only inconsistent with radiation-dominated epoch but
 also the energy conditions associated with the BD scalar
 field are violated. Therefore, the FLRW model with
 a varying BD coupling parameter has been studied.

By employing modified BD
models \cite{MM01,qiang2005,qiang2009,Ponce1,Ponce2,RFS11} constructed from
combining the standard BD setting with modern Kaluza-Klein
theory, it seems that some of the mentioned
problems (presented in the above paragraph) have been addressed.
Such modified frameworks, in which not only the matter but also a scalar potential emerge from the
geometry (see also \cite{RM18} and references therein), were inspired
by the space-time-matter theory (or the induced-matter
theory (IMT)) \cite{PW92,OW97,stm99}.
Inspired by \cite{RRT95}, the MBDT has also been extended to
arbitrary dimensions (for higher-dimensional cosmological
models, see, for instance, \cite{GGCC04,GC07,LP10,TSYA19}).
Moreover, the IMT and modified scalar-tensor theories
have been applied to establish interesting cosmological models \cite{RM18,RFM14,RPSM20,DRJ09,RJ10}.


In what follows, let us briefly present the objectives of this paper.

In Section \ref{OT-solution}, by considering the BD
theory and the spatially flat
FLRW universe filled with a perfect fluid in a $D$-dimensional spacetime,
we will retrieve the GDE in arbitrary dimensions.
Then, by restricting our attention to the fundamental
observers and past directed null vector fields, we will obtain Raychaudhuri
equation and the observer area distance.
Moreover, we will show that the Ricci focusing condition depends not
only the ordinary matter fields but also on those associated with the BD scalar field.

 In Section \ref{GR}, assuming that the BD scalar field takes constant values and ordinary matter has
 contribution from both dust and radiation, we will apply the
 formalism obtained in Section \ref{OT-solution} for a few
 particular cases in the context of GR.

 In Section \ref{D-dim BD}, we will study the BD cosmology (for which
 we establish two settings for the corresponding dynamical system in \ref{dynamical}).
 We also retrieve the energy conditions (i.e., null energy condition (NEC), weak
 energy condition (WEC), strong
  energy condition (SEC) and dominant energy condition (DEC)) associated with the BD
 cosmology in arbitrary dimensions and then check them for our exact solutions.
 Subsequently, concerning the well-known cosmological models, not
 only we will obtain extended exact solutions in arbitrary
 dimensions but also we will add corrections and modifications to those studied before.
 Then, the behavior of the deviation
 vector as well as observer area distance will be depicted for each model.

  In Section \ref{MBDT}, we will present a brief review of the MBDT
  framework. We will correct the definition of the induced scalar
  potential used in \cite{RFM14} (because for $D\neq4$, it
  makes the wave equation to be inconsistent with other field equations).
  Subsequently, we will show that the formalism of
  the GDE obtained in Section \ref{OT-solution} can also be employed for this
  framework. In order to apply the GDE for the
  cosmological models established in the MBDT, we would employ new
  procedures to investigate the
  cosmological solutions obtained in \cite{RFM14} and correct the wrong parts.
    We will argue that the cosmological solutions obtained in the MBDT framework have a few
  advantages with respect to the corresponding ones retrieved in the context
  of the conventional BD theory.
  Finally, we compare the behavior of the quantities of all models presented in this paper.
  In the last section, we will provide a summary and present further discussions.

\section{GDE in BD theory in $D$ dimensions}
\label{OT-solution}

\indent

In this section, let us first present a brief review of the BD
theory in the presence of a scalar potential in arbitrary dimensions.
Then, we obtain the GDE associated with our herein model
with a FLRW background.


In the Jordan frame, the action of the BD theory in $D$-dimensions can be written as
 \begin{equation}\label{D-action}
{\cal S}^{^{(D)}}=\int d^{^{\,D}}\!x \sqrt{-g}\,\left[\phi
R^{^{(D)}}-\frac{\omega}{\phi}\, g^{\alpha\beta}\,(\nabla_\alpha\phi)(\nabla_\beta\phi)-V(\phi)+\,
L\!^{^{(D)}}_{_{\rm matt}}\right],
\end{equation}
where $\phi$ and $\omega$ are the BD scalar field and the
dimensionless BD coupling parameter\rlap,\footnote{
In a $D$-dimensional
spacetime, imposing the constraint
$\omega>\!-(D-1)/(D-2)$ enables us to transform the BD
theory from the Jordan frame to the Einstein frame.} respectively. In
this paper, we have chosen the units $c=1=8\pi G_0$ (where $c$ and $G_0$ stand
for the speed of light and the Newton gravitational constant, respectively).
Therefore, the present value of the BD scalar field (which is normalized to $G_0^{-1}$) should
be equal to unity, see \cite{RMM19} and references therein.
 The Greek indices run from zero
to $D-1$; $g$ and $R^{^{(D)}}$ denote respectively the determinant
and Ricci curvature scalar of the $D$-dimensional
spacetime metric $g_{\alpha\beta}$; $\nabla_\alpha$
denotes the covariant derivative in a $D$-dimensional
spacetime; $L^{^{(D)}}_{_{\rm matt}}$ is the
Lagrangian associated with the ordinary matter.


Variation of the action \eqref{D-action} yields
  \begin{eqnarray}\label{BD-Eq-DD}
G_{\mu\nu}^{^{(D)}}=\frac{T_{\mu\nu}}{\phi}+
\frac{\omega}{\phi^2}\left[\left(\nabla_\mu\phi\right)\left(\nabla_\nu\phi\right)-
\frac{1}{2}g_{\mu\nu}(\nabla_\alpha\phi)(\nabla^\alpha\phi)\right]
+\frac{1}{\phi}\left(\nabla_\mu\nabla_\nu\phi- g_{\mu\nu}\nabla^2\phi\right)
-g_{\mu\nu}\frac{V\left(\phi\right)}{2\phi},
\end{eqnarray}
\begin{eqnarray}\label{D2-phi}
\nabla^2\phi=\frac{1}{\chi(D)}
\left[T+\left(\frac{D-2}{2}\right)\phi\,\frac{dV\left(\phi\right)}{d\phi}
-\frac{D}{2}V\left(\phi\right)\right],
\end{eqnarray}
where $\nabla^2\equiv\nabla_\alpha\nabla^\alpha$ and
\begin{eqnarray}
\label{chi}
\chi(D)&\equiv&(D-2)\omega+(D-1).
\end{eqnarray}


A contraction of the equation \eqref{BD-Eq-DD} yields
\begin{eqnarray}\label{gen-Ricc-scalar}
R^{^{(D)}}=-\frac{2}{(D-2)\phi}\,T+
\frac{\omega}{\phi^2}(\nabla_\alpha\phi)(\nabla^\alpha\phi)
+\frac{1}{(D-2){\phi}}\left[2(D-1)\nabla^2\phi+D V(\phi)\right].
\end{eqnarray}
By substituting $R^{^{(D)}}$ from \eqref{gen-Ricc-scalar} to \eqref{BD-Eq-DD}, we obtain
\begin{eqnarray}\label{gen-Ricc-tensor}
R_{\mu\nu}^{^{(D)}}\!=\frac{1}{\phi}\left[T_{\mu\nu}-\frac{T}{(D-2)}g_{\mu\nu}\right]+
\frac{\omega}{\phi^2}(\nabla_\mu\phi)(\nabla_\nu\phi)
+\frac{1}{(D-2)\phi}\left[(D-2)\nabla_\mu\nabla_\nu\phi+g_{\mu\nu}\nabla^2\phi+g_{\mu\nu}V(\phi)\right].
\end{eqnarray}

In what follows, we first obtain the GDE for the herein BD
setting and then focus on the special cases in the subsections.
Let us suppose a congruence of geodesics with affine
 parameter $\zeta$ whose normalized tangent vector field $v^\alpha$ is defined as
 \begin{eqnarray}\label{t-vector}
v^\alpha\equiv
\frac{dx^\alpha(\zeta)}{d\zeta},
\end{eqnarray}
which satisfies \cite{EE97}
\begin{eqnarray}\label{t-vector2}
v_\alpha v^\alpha=\varepsilon, \hspace{10mm} \frac{{\cal D}v^\alpha}{{\cal D}\zeta}=v^\beta \nabla_\beta v^\alpha=0,
\end{eqnarray}
where $\frac{{\cal D}}{{\cal D}\zeta}$ denotes
the covariant derivation along the geodesics; $\varepsilon=-1,0,1$, if the
geodesics are timelike, null and spacelike, respectively.
Moreover, by considering a family of interpolating geodesics with affine parameter $\varsigma$,
we define the deviation (connecting) vector as
$\eta^\alpha\equiv \frac{dx^\alpha(\varsigma)}{d\varsigma}$, which
connects two of the neighboring geodesics in the congruence.
Therefore, $\eta^\alpha$ commutes with $v^\alpha$. In order to
simplify the corresponding equations, we can further assume \cite{EE97}:
\begin{eqnarray}\label{t-vector3}
\eta_\alpha v^\alpha=0.
\end{eqnarray}

The general GDE is given by\footnote{From now on, let us drop the
superscript $(D)$ of the quantities associated with a $D$-dimensional spacetime.}
\begin{eqnarray}\label{GDE-gen}
\frac{{\cal D}^2\eta^\alpha}{{\cal D}\zeta^2}=-R^\alpha_{\,\,\beta\gamma\delta}v^\beta \eta^\gamma v^\delta,
\end{eqnarray}
\\
where $R^\alpha_{\,\,\beta\gamma\delta}$ is the Riemann
tensor.
In $D$ dimensions, the Weyl
tensor $C_{\alpha\beta\gamma\delta}$ is related to the metric, Riemann
tensor, Ricci tensor and the Ricci scalar as \cite{Inverno.book}
\begin{eqnarray}\label{Rie-ten}
R_{\alpha\beta\gamma\delta}\!\!&=&\!\!C_{\alpha\beta\gamma\delta}
+\frac{1}{D-2}\left(g_{\alpha\gamma}R_{\delta\beta}-g_{\alpha\delta}
R_{\gamma\beta}+g_{\beta\delta}R_{\gamma\alpha}-g_{\beta\gamma}R_{\delta\alpha}\right)\\\nonumber
\!\!&+&\!\!\frac{R}{(D-1)(D-2)}\left(g_{\alpha\gamma}g_{\delta\beta}
-g_{\alpha\delta}g_{\gamma\beta}\right),\hspace{8mm} {\rm for}\hspace{8mm}D\geq3.
\end{eqnarray}

Now, let us restrict our attention to
a $D$-dimensional spacetime described by the spatially flat FLRW
metric:
\begin{equation}\label{brane-metric}
ds^2=-dt^{2}+a^{2}(t)\left[\sum^{D-1}_{i=1}
\left(dx^{i}\right)^{2}\right],
\end{equation}
where $t$ and $x^i$ are the cosmic time and the
Cartesian coordinates, respectively; $a(t)$ is the scale
factor. We should mention that, due to the spacetime
symmetries, the components of the FLRW metric as well
as the scalar field $\phi$ should depend only on the
comoving time.

Moreover, we shall restrict ourselves to a perfect fluid whose energy-momentum tensor is given by
 \begin{eqnarray}\label{perfect}
T_{\mu\nu}=\left(\rho+p\right)u_\mu u_\nu+pg_{\mu\nu},
\end{eqnarray}
where the $D$-vector velocity of the fluid satisfies $u^\mu u_\mu=-1$.
Therefore, the trace of the energy-momentum tensor in $D$ dimensions is given by
\begin{eqnarray}\label{perfect-trace}
T=-\rho+\left(D-1\right)p.
\end{eqnarray}

Concerning computing the GDE in the context of the BD theory,
relations \eqref{perfect} and \eqref{perfect-trace} will be useful.
In the following four steps, we briefly describe how we can compute the GDE for our herein model.
(i) The Weyl tensor associated with the FLRW metric vanishes. Therefore, equation \eqref{Rie-ten} yields
  \begin{eqnarray}\nonumber
R^\lambda_{\,\,\beta\gamma\delta}v^\beta \eta^\gamma v^\delta=
\frac{1}{D-2}\left[\left(\delta^\lambda_{\gamma}R_{\delta\beta}-\delta^\lambda_\delta
R_{\gamma\beta}+g^{\lambda\alpha}g_{\beta\delta}R_{\gamma\alpha}-g^{\lambda\alpha}
g_{\beta\gamma}R_{\delta\alpha}\right)
-\frac{R}{(D-1)}\left(\delta^\lambda_\gamma g_{\delta\beta}-\delta^\lambda_\delta
g_{\gamma\beta}\right)\right]v^\beta \eta^\gamma v^\delta.\\\nonumber\\
\label{Rie-ten-2}
\end{eqnarray}
\\
(ii)
    We substitute
    the corresponding components of the energy-momentum tensor and its trace from
    relations \eqref{perfect} and \eqref{perfect-trace} into equation \eqref{gen-Ricc-tensor}.\\
(iii) Subsequently, we can easily derive the Ricci scalar from the Ricci tensor obtained in (ii).\\
(iv) We substitute the corresponding components of the Ricci tensor and
   Ricci scalar from steps (ii) and (iii) to
    equation \eqref{Rie-ten-2} and then
    employ $v_\alpha v^\alpha=\varepsilon$, $-v_\alpha u^\alpha=E$
    and $\eta_\alpha v^\alpha=0=\eta_\alpha u^\alpha$ (to read a complete
    explanations concerning these conditions see, for instance, \cite{EE97}).\\
Following the steps presented above, it is
 straightforward to show that the GDE in the BD theory in $D$ dimensions reads
\begin{eqnarray}
\frac{{\cal D}^2\eta^\lambda}{{\cal D}\zeta^2}=-R^\lambda_{\,\,\beta\gamma\delta}v^\beta \eta^\gamma v^\delta
=-\left\{E^2\left[\frac{\rho_{\rm tot}+p_{\rm tot}}{(D-2)\phi}\right]
+2\varepsilon \left[\frac{\rho_{\rm tot}}{(D-1)(D-2)\phi}\right]\right\}\eta^\lambda,
\label{GDE-gen-2}
\end{eqnarray}
In equation \eqref{GDE-gen-2}, the quantities $\rho_{\rm tot}$ and $p_{\rm tot}$
stand, respectively, for the sum of the energy density and pressure associated
with both the ordinary matter and the BD scalar field. Namely,
\begin{eqnarray}
 \rho_{\rm tot}\equiv\rho+\rho_\phi, \hspace{10mm}
 p_{\rm tot}\equiv p+p_\phi.
 \label{ro-pi-total}
\end{eqnarray}
Moreover, in analogy with the BD cosmology in four dimensions (see~\cite{T02} and references therein), we have
defined the energy density and pressure associated with the BD scalar field in $D$ dimensions, respectively, as
\begin{eqnarray}
\label{rho-phi-gen}
\rho_\phi\!\!&\equiv\!\!&\frac{\omega}{2}
\frac{\dot{\phi}^2}{\phi}+\frac{V(\phi)}{2}-(D-1)H\dot{\phi},
\\\nonumber\\
p_\phi\!\!&\equiv\!\!&\frac{\omega}{2}
\frac{\dot{\phi}^2}{\phi}-\frac{V(\phi)}{2}+\ddot{\phi}+(D-2)H\dot{\phi},
\label{p-phi-gen}
\end{eqnarray}
where $H\equiv \dot{a}/a$ is the Hubble parameter.
Equation \eqref{GDE-gen-2} is the $D$-dimensional GDE corresponding to the FLRW background in the
 context of the BD theory. Moreover, in the particular cases
 where $D=4$, $\phi={\rm constant}$ and $V(\phi)=0$,
 equation \eqref{GDE-gen-2} reduces to the Pirani equation \cite{S34,P56,EE97}.

\subsection{GDE for fundamental observers}
\label{fundamental}
For this particular case, $v^\alpha=u^\alpha$, therefore from \eqref{t-vector2}, we get $\varepsilon=-1$, and for a normalized vector field,
 we have $E=1$. Moreover, $\zeta$ coincides with $t$ \cite{EE97}. Therefore, \eqref{GDE-gen-2} reduces to
\begin{eqnarray}
R^\lambda_{\,\,\beta\gamma\delta}u^\beta \eta^\gamma u^\delta=
\left[\frac{ (D-3) \rho_{\rm tot}+(D-1) p_{\rm tot}}{(D-1)(D-2)\phi}\right]\eta^\lambda.
\label{fun-0}
\end{eqnarray}
Assuming the deviation vector as $\eta_\lambda=\vartheta e_\lambda$ (where $ e_\lambda$
is parallel propagated along the cosmic time), isotropy leads to have
\begin{eqnarray}
\frac{de^\lambda}{dt}=0.
\label{fun-1}
\end{eqnarray}
Therefore, we get
\begin{eqnarray}
\frac{d^2\eta^\lambda}{dt^2}=\frac{d^2\vartheta}{dt^2}e^\lambda.
\label{fun-2}
\end{eqnarray}
 Consequently, equation \eqref{GDE-gen-2} reduces to
\begin{eqnarray}
\frac{d^2{\vartheta}}{dt^2}=-\left[\frac{(D-3) \rho_{\rm tot}+(D-1) p_{\rm tot}}{(D-1)(D-2)\phi}\right]{\vartheta}.
\label{fun-3}
\end{eqnarray}
Specifically for $\vartheta=a(t)$, equation \eqref{fun-3} reads
\begin{eqnarray}\label{fun-Ray}
\frac{\ddot{a}}{a}=-\left[\frac{(D-3) \rho_{\rm tot}+(D-1) p_{\rm tot}}{(D-1)(D-2)\phi}\right],
\end{eqnarray}
which is the Raychaudhuri equation corresponding to the spatially flat
FLRW metric in the context of the BD theory in $D$-dimensions.

Substituting $\rho_\phi$ and $p_\phi$ from relations \eqref{rho-phi-gen}
and \eqref{p-phi-gen} into equation \eqref{fun-Ray}, the
Raychaudhuri equation is rewritten as
\begin{eqnarray}
\frac{\ddot{a}}{a}=-\frac{(D-3)\rho+(D-1)p}{(D-1)(D-2)\phi}
-\frac{\omega}{(D-1)}\left(\frac{ \dot{\phi}}{\phi}\right)^2-\frac{1}{(D-2)}\left[\frac{\ddot{\phi}}{\phi}
+H\left(\frac{\dot{\phi}}{\phi}\right)\right]+\frac{V({\phi})}{(D-1)(D-2)\phi}.
\label{fun-Ray-1}
\end{eqnarray}


It is straightforward to show that equation \eqref{fun-Ray} can also be
derived from the following field equations associated with the BD
cosmology when the universe is described with the spatially flat
FLRW line-element \cite{RFM14}:
\begin{eqnarray}
\label{fun-Fri-1}
\frac{(D-1)(D-2)}{2}H^2
\!\!&=&\!\!\frac{ \rho_{\rm tot}}{\phi},\\\nonumber\\
\label{fun-Fri-2}
(D-2)\frac{\ddot{a}}{a}+\frac{(D-2)(D-3)}{2}H^2
\!\!&=&\!\!-\frac{ p_{\rm tot}}{\phi},
\end{eqnarray}
where it was assumed that the ordinary matter being a
perfect fluid which satisfies the conservation law:
\begin{eqnarray}
\label{BD-FRW4}
\dot{\rho}+(D-1) H (p+\rho )=0.
\end{eqnarray}
However, using the above assumption as well as the Bianchi identity, it
is easy to show that $\rho_{\phi}$ and $\rho_{\phi}$ (which are given by
relations \eqref{rho-phi-gen} and \eqref{p-phi-gen}) do not satisfy such a conservation law:
\begin{eqnarray}
\label{cons-phi-matt}
\dot{\rho}_\phi+(D-1) H (p_\phi+\rho_\phi)\neq0.
\end{eqnarray}

Substituting dimensionless cosmological density parameters
\begin{eqnarray}
\label{density.par.def}
\Omega\equiv\frac{2}{(D-1)(D-2)}\frac{\rho}{H^2\phi},\hspace{10mm}
\Omega_{\phi}\equiv\frac{2}{(D-1)(D-2)}\frac{\rho_{\phi}}{H^2\phi},
\end{eqnarray}
into equation \eqref{fun-Fri-1}, we obtain
\begin{eqnarray}
\label{fri-eq}
\Omega+\Omega_{\phi}=1.
\end{eqnarray}

Moreover, the wave equation \eqref{D2-phi} corresponding to
the $D$-dimensional flat FLRW \eqref{brane-metric} is:
\begin{eqnarray}
\label{BD-FRW3}
\ddot{\phi}+(D-1) H \dot{\phi}=
\frac{1}{\chi(D)}\left[\rho -(D-1) p+\frac{D}{2} V(\phi)
-\left(\frac{D-2}{2}\right)\phi\,\frac{dV(\phi)}{d\phi} \right].
\end{eqnarray}
We should note that among the four FLRW-BD field equations
\eqref{fun-Fri-1}-\eqref{BD-FRW4} and \eqref{BD-FRW3}, only three of them are independent.

\subsection{GDE for null vector fields}

In this case, $v^\alpha=k^\alpha$ such that $k_\alpha k^\alpha=0$ \cite{EE97}.
Therefore, by substituting $\varepsilon=0$ to equation \eqref{GDE-gen-2}, we obtain
\begin{eqnarray}
R^\lambda_{\,\,\beta\gamma\delta}k^\beta \eta^\gamma k^\delta=
E^2\left[\frac{  \rho_{\rm tot}+ p_{\rm tot}}{(D-2)\phi}\right]\eta^\lambda.
\label{BD-R.foc-null}
\end{eqnarray}

Moreover, setting $\eta^\lambda=\eta e^\lambda$, $e_\lambda e^\lambda=1$,
$e_\lambda u^\lambda=e_\lambda k^\lambda=0$ and employing a parallelly propagated and aligned basis, i.e.,
$\frac{{\cal D}e^\lambda}{{\cal D}\zeta}=k^\alpha \nabla_\alpha e^\lambda=0$
\cite{EE97}, equation \eqref{GDE-gen-2} reduces to
\begin{eqnarray}
\frac{d^2\eta}{d\zeta^2}=-E^2\left[\frac{  \rho_{\rm tot}+ p_{\rm tot}}{(D-2)\phi}\right]\eta.
\label{BD-null}
\end{eqnarray}
Equation \eqref{BD-null} yields the Ricci
focusing in $D$ dimensions as
\begin{eqnarray}
\frac{ \rho_{\rm tot}+ p_{\rm tot}}{(D-2)\phi}>0.
\label{BD-foc}
\end{eqnarray}
We should note that in order to obtain Ricci focusing condition in the BD
theory as well as to recover the corresponding GR case, we can admit the following
requirement: from action \eqref{D-action}, it can be read that
the gravitational coupling is \cite{Faraoni.book}
\begin{eqnarray}
G_{\rm eff}(\phi)=\frac{1}{\phi},
\label{G-coupling}
\end{eqnarray}
which implies that the BD scalar field should take positive values for an attractive gravity.
For the particular case where $\phi\sim\frac{1}{G_0}=1$, equation \eqref{BD-null} reduces to
\begin{eqnarray}
\frac{d^2\eta}{d\zeta^2}=-E^2\left(\frac{\rho+p}{D-2}\right)\eta,
\label{GR-null}
\end{eqnarray}
which is a generalization of that obtained
in \cite{EE97} in four dimensions where the focusing condition
for all families of the past-directed as well as future-directed null
geodesics is given by $\rho+p>0$; while for a fluid with a
equation of state (EoS) $W=-1$ (cosmological constant)
 there is not any influence in the focusing \cite{EE97}.

For the particular case of the herein BD model corresponding to de Sitter (like)
universe, i.e., choosing $ p+\rho=0$, equation \eqref{BD-null} is simplified:
\begin{eqnarray}
\frac{d^2\eta}{d\zeta^2}=-E^2\left[\frac{\rho_{\phi}+p_{\phi}}{(D-2)\phi}\right]\eta
=-\frac{E^2}{D-2}\left[\frac{\ddot{\phi}}{\phi}+\omega
\left(\frac{\dot{\phi}}{\phi} \right)^2-H \left(\frac{\dot{\phi}}{\phi }\right)
\right]\eta.
\label{BD-null-DeS}
\end{eqnarray}

It is worth expressing equation \eqref{BD-null} in terms of the redshift parameter $z$.
For the null geodesics, we have \cite{EE97}
\begin{eqnarray}
1+z=\frac{a_0}{a}=\frac{E}{E_0},
\label{redshift}
\end{eqnarray}
where $a_0=1$ is the present value of the scale factor.
Moreover, the differential operators are transformed as
\begin{eqnarray}\label{ops1}
\frac{d}{d\zeta}&=&\frac{dz}{d\zeta}\frac{d}{dz},\\\nonumber
\\
\frac{d^2}{d\zeta^2}&=&\left(\frac{d\zeta}{dz}\right)^{-2}
\left[\frac{d^2}{dz^2}-\left(\frac{d\zeta}{dz}\right)^{-1}\frac{d^2\zeta}{dz^2}\frac{d}{dz}\right].
\label{ops2}
\end{eqnarray}
Therefore, for the past directed case, using
\begin{eqnarray}\label{ops3-2}
\frac{dt}{d\zeta}=E=E_0(1+z),
\end{eqnarray}
and \eqref{redshift}, we easily obtain
\begin{eqnarray}\label{ops3}
\frac{d\zeta}{dz}&=&\frac{1}{E_0H(1+z)^2},\\\nonumber\\
\frac{d^2\zeta}{dz^2}&=&\frac{1}{E_0H^3(1+z)^3}\left(\frac{\ddot{a}}{a}-3H^2\right).
\label{ops4}
\end{eqnarray}
To obtain equation \eqref{ops4}, we have substituted $dH/dz$ as
\begin{eqnarray}\label{ops3-1}
\frac{dH}{dz}=\frac{d\zeta}{dz}\frac{dt}{d\zeta}\frac{dH}{dt}=-\frac{\dot{H}}{H(1+z)},
\end{eqnarray}
which, in turn, obtained from using \eqref{ops3-2}.

Now, we can compute equation \eqref{ops2} for $\eta$:
we substitute ${d^2\eta}/{d\zeta^2}$, ${d\zeta}/{dz}$
and ${d^2\zeta}/{dz^2}$, respectively,
from \eqref{BD-null}, \eqref{ops3} and \eqref{ops4}
into \eqref{ops2} and then substituting $\ddot{a}/a$
from equation \eqref{fun-Ray}; we easily obtain:
\begin{eqnarray}\label{ops5}
\frac{d^2\eta}{dz^2}+\frac{1}{1+z}\left[3+\frac{(D-3) \rho_{\rm tot}
+(D-1) p_{\rm tot}}{(D-1)(D-2)H^2\phi}\right]\frac{d\eta}{dz}
+\left[\frac{ \rho_{\rm tot}+ p_{\rm tot}}{(D-2)(1+z)^2H^2\phi}\right]\eta=0.
\end{eqnarray}

We should note that using definitions \eqref{density.par.def} and equation \eqref{fri-eq},
it is straightforward to show that equation \eqref{ops5} can also be written as
\begin{eqnarray}\label{ops5-1}
\frac{d^2\eta}{dz^2}+\frac{1}{1+z}
\left[\frac{D+3}{2}+\frac{p_{\rm tot}}{(D-2)H^2\phi}\right]
\frac{d\eta}{dz}
+\frac{1}{(1+z)^2}\left[\frac{D-1}{2}+\frac{p_{\rm tot}}{(D-2)H^2\phi}\right]\eta=0.
\end{eqnarray}

Equation \eqref{ops5} (or equivalently \eqref{ops5-1})
 is the general GDE associated with the null vector fields with the FLRW
 background in the context of the BD framework (with a non-vanishing scalar potential),
 which is also applicable for the MBDT, cf Section \ref{MBDT}.

 It is worthwhile to obtain a general solution for equation \eqref{ops5-1}.
 Assuming that the expressions in the brackets in \eqref{ops5-1} being two general
 functions of the redshift parameter
 and defining the new variables
 \begin{eqnarray}\label{ops5-1-sol-1}
 \tilde{x}\equiv 1+z,\hspace{10mm} \eta\equiv \frac{\tilde{y}}{\tilde{x}}, \hspace{10mm}
 \ell(\tilde{x})\equiv \frac{D+3}{2}+\frac{p_{\rm tot}}{(D-2)H^2\phi},
\end{eqnarray}
equation \eqref{ops5-1} is written as
\begin{eqnarray}\label{ops5-1-sol-2}
\frac{d^2\tilde{y}}{d\tilde{x}^2}+\left[\frac{\ell(\tilde{x})-2}{\tilde{x}}\right]\frac{d\tilde{y}}{d\tilde{x}}=0.
\end{eqnarray}
It is straightforward to show that \eqref{ops5-1-sol-2} yields a
general solution in terms of the redshift parameter as
\begin{eqnarray}\label{ops5-1-sol-3}
\eta(z)=\frac{\eta_1}{1+z}+\frac{\eta_2}{1+z}\int dz
\left\{Exp\left[\int^z dz'\left(\frac{D-1}{2}+\frac{ p_{\rm tot}}{(D-2)H^2\phi}\right)\right]\right\},
\end{eqnarray}
where $\eta_1$ and $\eta_2$ are constants of integration.

 In the next section, we
 first focus on the GR limit of the model and then in sections
 \ref{D-dim BD} and \ref{MBDT} we will proceed our discussions for a
 few interesting cosmological models in BD theory and in the MBDT, respectively.
 We should note that for the later, we have to use the components of the energy-momentum tensor
 as well as the scalar potential, which are dictated from the geometry (see Section \ref{MBDT}).

\section{GDE for null vector field in GR}
\label{GR}

In order to retrieve the GDE in GR limit in $D$-dimensions, let us first rewrite
the general formalism of the GDE for the null vector fields in the BD model
by assuming a particular perfect fluid.
Namely, letting the energy density and the pressure of the ordinary
matter have contributions from both incoherent matter and radiation:
  \begin{eqnarray}\label{matt-rad-1}
\rho \!&=&\!(D-1)H_0^2(1+z)^{D-1}\Big[\Omega_{m0}+\Omega_{r0}(1+z)\Big],\\\nonumber\\
\label{matt-rad-2}
p\!&=&\!H_0^2\Omega_{r0}(1+z)^{D},
\end{eqnarray}
 where we have assumed $p=p_r=\rho_r/(D-1)$ (where the index ${\rm r}$
 refers to the radiative fluid) for which the conservation law is also satisfied.
Therefore, the Friedmann equation \eqref{fun-Fri-1} is rewritten as
\begin{eqnarray}
\label{fun-Fri-3}
H^2
\!\!&=&\!\!H_0^2\left\{\frac{2}{(D-2)\phi}\left[\Omega_{m0}(1+z)^{D-1}
+\Omega_{r0}(1+z)^D\right]+\Omega_{\rm DE}\right\},
\end{eqnarray}
where
\begin{eqnarray}
\label{DE-omega}
\Omega_{\rm DE}\equiv \frac{2\,\rho_{\phi}}{(D-1)(D-2)H_0^2\phi}.
\end{eqnarray}

For this case, equation \eqref{ops5} reduces to
\begin{eqnarray}\label{ops6}
\frac{d^2\eta}{dz^2}+{\cal P}\frac{d\eta}{dz}+{\cal Q}\eta=0,
\end{eqnarray}
where we defined ${\cal P}={\cal P} (H,dH/dz,z,D)$ and ${\cal Q}={\cal Q} (H,dH/dz,z,D)$ as
\begin{eqnarray}\label{P}
{\cal P}\!\!&\equiv&\!\!\frac{3}{(1+z)}\\\nonumber
&+&\frac{(D-1)H_0^2(1+z)^{D-1}\left[(D-3)\Omega_{m0}+(D-2)\Omega_{r0}(1+z)\right]
+\left[(D-3)\rho_{\phi}+(D-1)p_{\phi}\right]}
{2(D-1)H_0^2(1+z)^{D}\left[\Omega_{m0}+\Omega_{r0}(1+z)\right]+2(1+z)\rho_{\phi}},\\\nonumber\\\nonumber\\
\label{Q}
{\cal Q}&\equiv& \frac{(D-1)}{2(1+z)^2}\,\Bigg\{\frac{H_0^2(1+z)^{D-1}\left[(D-1)\Omega_{m0}+D\Omega_{r0}(1+z)\right]
+\left(\rho_{\phi}+p_{\phi}\right)}
{(D-1)H_0^2(1+z)^{D-1}\left[\Omega_{m0}+\Omega_{r0}(1+z)\right]+\rho_{\phi}}\Bigg\}.
\end{eqnarray}

We should note that, with the
assumptions \eqref{matt-rad-1} and \eqref{matt-rad-2}, equations
\eqref{fun-Fri-3}-\eqref{Q} are completely general and can be used for the BD theory.

In the limit $\omega\rightarrow \infty$, the
standard BD theory including a scalar potential reduces
to $GR+\Lambda$ where $\Lambda$ is a constant \cite{T02}.
Let us investigate this particular case in our model:
setting
 $\phi\sim 1/G_0=1$
and $V\equiv2\Lambda$, relations \eqref{rho-phi-gen}
and \eqref{p-phi-gen} yield $\rho_{\phi}=-p_{\phi}=\Lambda$.
Therefore, equation \eqref{GDE-gen-2} reduces to
\begin{eqnarray}
\frac{{\cal D}^2\eta^\lambda}{{\cal D}\zeta^2}=-R^\lambda_{\,\,\beta\gamma\delta}v^\beta \eta^\gamma v^\delta
=-\Bigg\{E^2\left(\frac{ \rho+p}{D-2}\right)
+2\varepsilon \left[\frac{\rho+\Lambda}{(D-1)(D-2)}\right]\Bigg\}\eta^\lambda,
\label{GDE-gen-3}
\end{eqnarray}
which can be considered as a generalization of the Pirani equation \cite{S34,P56,EE97} to $D$-dimensions.
It is seen that for $D=4$, we
retrieve the same equation obtained in \cite{EE97}.

Moreover, for the particular conditions associated with the GR limit, equation \eqref{DE-omega} reduces to
\begin{eqnarray}\label{omeg-lam}
\Omega_{\rm DE}= \frac{2 \Lambda}{(D-1)(D-2)H_0^2}\equiv \Omega_\Lambda={\rm constant}.
\end{eqnarray}
Therefore, it is straightforward to show that equation \eqref{ops6} reduces to
\begin{eqnarray}\nonumber
\frac{d^2\eta}{dz^2}&+&\Bigg\{\frac{(1+z)^{D-1}\left[(D+3)\Omega_{m0}+(D+4)\Omega_{r0}(1+z)\right]+2(D-2)\Omega_\Lambda}
{2(1+z)^D\left[\Omega_{m0}+\Omega_{r0}(1+z)\right]+(D-2)(1+z)\Omega_\Lambda}\Bigg\}\,\frac{d\eta}{dz}\\\nonumber\\
&+&
\Bigg\{\frac{(D-1) \Omega _{m0}+D (1+z) \Omega _{r0}}{2 (1+z)^2
\left[\Omega _{m0}+(1+z) \Omega _{r0}\right]+(D-2) (1+z)^{3-D} \Omega _{\Lambda }}\Bigg\}\,\eta=0. \label{GR-D-CC}
\end{eqnarray}
Equation \eqref{GR-D-CC} is
the GDE for the null vector fields associated
with the $GR+\Lambda$ framework in $D$ dimensions.
Let us investigate the solutions associated with \eqref{GR-D-CC} for the following cases.

\begin{figure}
\centering{}\includegraphics[width=2.6in]{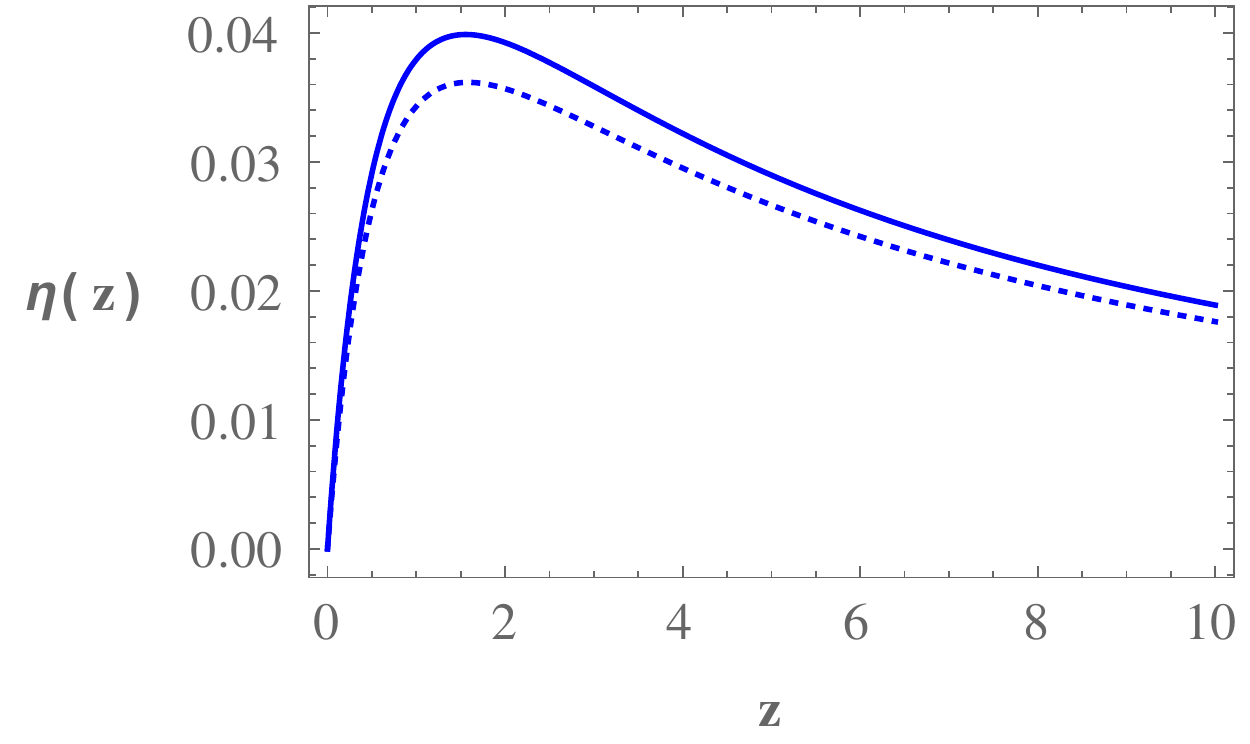}\hspace{2mm}
\includegraphics[width=2.6in]{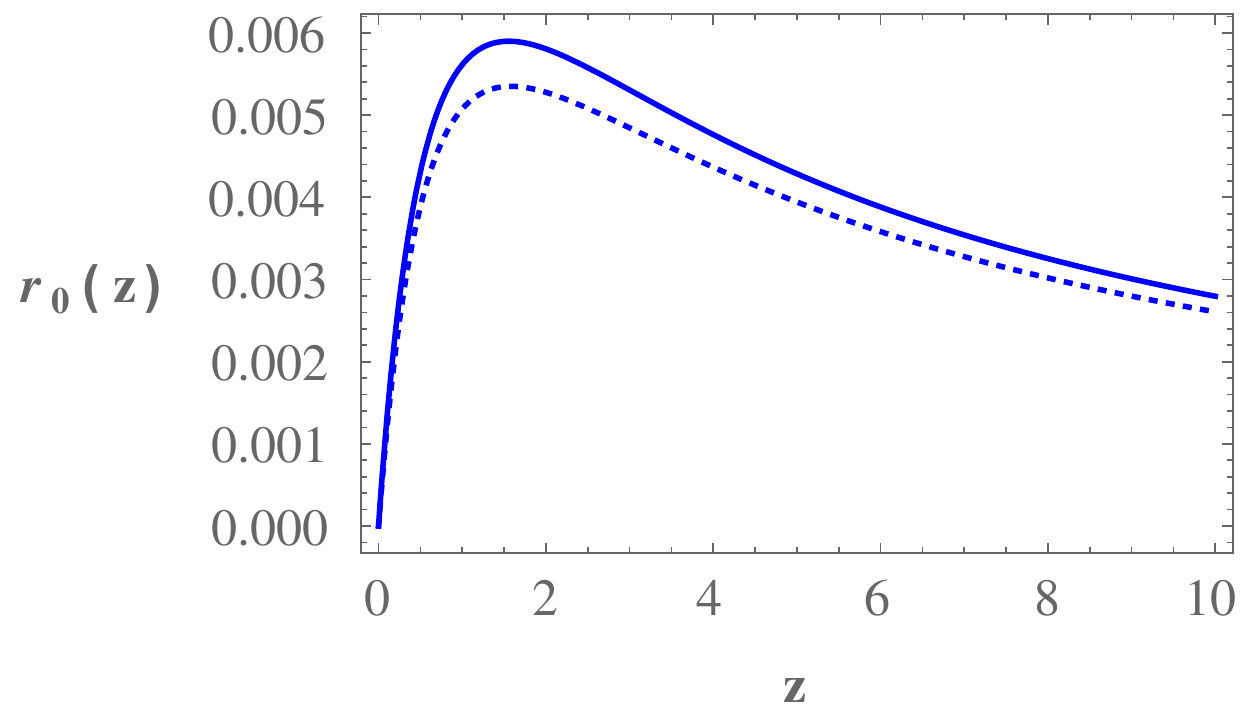}
\caption{{\footnotesize The behavior
of the deviation vector $\eta(z)$ (the left panel) and observer area distance
$r_0(z)$ (the right panel) in the units of $H_0^{-1}$ for null vector fields
with FLRW background in GR in the presence of the cosmological constant.
We have assumed $D=4$, $H_0=67.6 km/s/Mpc$,
$\eta(0)=0$, $\frac{d\eta(z)}{dz}\mid_{z=0}=0.1$, $\Omega_{\Lambda}=0.68$
and $8\pi G_0=1$.
For plotting the solid and the dotted curves, we have assumed
$\{\Omega_{m0}=0.31, \Omega_{r0}=0.01\}$ and  $\{\Omega_{m0}=0.32,\Omega_{r}=0\}$,
respectively, without changing the other initial conditions and parameters.
For more visibility, the dotted curves were re-scaled.}}
\label{eta-r-GR-DD}
\end{figure}

(i) For the most generalized case where $\Lambda\neq0$, $\Omega_{m0}\neq0$ and $\Omega_{r0}\neq0$,
let us investigate the solution for the differential equation \eqref{GR-D-CC} by applying numerical codes.
In this respect, we have employed a numerical code to depict the
behavior of the deviation vector in
terms of the redshift parameter in the presence of
the cosmological constant, see the left panel of the
figure \ref{eta-r-GR-DD}. (Throughout this paper, we will use the observational
data reported in \cite{PL.CosPar18}.) Although, here we have
prepared the figures only for $D=4$, but our numerical endeavors
 have shown that the behavior of $\eta(z)$ is affected by the number of the dimensions of spacetime.

(ii) For a particular case where $\Lambda=0$, the analytic
solution of equation \eqref{GR-D-CC} is given by a complicated
function as\footnote{To the best of our knowledge, both of
the cases (i) and (ii) (concretely, only for $D\neq4$ associated with the case (ii)) have not been investigated in GR.}
\begin{eqnarray}\label{GR-D-NOCC}
\eta (z)=\frac{C_1}{(z+1)}+ \frac{2 C_2  \left(-\frac{\Omega_{r0}}{\Omega_{m0}}\right)^{\frac{D-1}{2}}
\left(\Omega_{m0}+\Omega_{r0}+\Omega_{r0} z\right)^{\frac{1}{2}} \, _2F_1\left(\frac{1}{2},\frac{D-1}{2};\frac{3}{2};
\frac{\Omega_{m0}+z \Omega_{r0}+\Omega_{r0}}{\Omega_{m0}}\right)}{\Omega_{r0}(z+1)}, \end{eqnarray}
where the integration constants $C_1$ and $C_2$ carry the
dimension of $\eta(z)$ and $_2F_1(a,b;c;z)$ is the hypergeometric function.

Moreover, for the special case where $D=4$, the solution \eqref{GR-D-NOCC} reduces to
\begin{eqnarray}\label{GR-4-NOCC}
\eta (z)=\frac{C_1}{z+1}-\frac{2 C_2 \left(\Omega_{m0}
+\Omega_{r0}+\Omega_{r0} z\right)^{\frac{1}{2}}}{\Omega_{m0} (z+1)^{\frac{3}{2}}}.
\end{eqnarray}
The solution \eqref{GR-4-NOCC}, by re-scaling the integration
constants, has also been obtained in \cite{EE97}.

The relation for the observer area distance $r_0(z)$ is given by
 \begin{eqnarray}\label{mattig-1}
r_0(z)=\sqrt{\Bigl|{}\frac{dA_0(z)}{d\Omega_s}\Bigr|{}}=
\Bigl|{}\frac{\eta(z')\mid_z}{d\eta(z')/d\ell\mid_{z=0}}\Bigr|{},
\end{eqnarray}
where $A_0$ and $\Omega_s$ stand for the area of the object and the
solid angle, respectively. In order to obtain $r_0(z)$ we should
employ $d/d\ell=E_0^{-1}(1+z)^{-1}d/d\zeta=H(1+z)d/dz$ and assume
that the integration constants for $\eta(z)$ are related to each
other via $\eta(z=0)=0$. Obviously, without having an exact
solution for the deviation vector, it is not possible to obtain an
exact expression for $r_0(z)$.

For the case (i), let us
focus on numerical approaches to depict the behavior of $r_0(z)$.
In the right panel of figure \ref{eta-r-GR-DD}, we have plotted the
behavior of $r_0(z)$ for $D=4$, see the solid curve.

For the case (ii), using \eqref{GR-D-NOCC}, we obtain
\begin{eqnarray}\nonumber
r_0(z)&=& \frac{2(\Omega_{m0}+\Omega_{r0}) \,\, _2F_1\left(1,\frac{4-D}{2};\frac{3}{2};
\frac{\Omega_{m0}+\Omega_{r0}}{\Omega_{m0}}\right)}{{H_0 \Omega_{m0}}(z+1)}\\\nonumber
\\
&-&\frac{2
\Big[(\Omega_{m0}+\Omega_{r0})(\Omega_{m0}+\Omega_{r0}
+\Omega_{r0} z)\Big]^{\frac{1}{2}} \, _2F_1\left(1,\frac{4-D}{2};\frac{3}{2};
\frac{\Omega_{m0}+\Omega_{r0}(z+1)}{\Omega_{m0}}\right)}
{{H_0 \Omega_{m0}}(z+1)^{\frac{D-1}{2}}}.\label{mattig-GD-D-0}
\end{eqnarray}

\begin{figure}
\centering{}\includegraphics[width=3.2in]{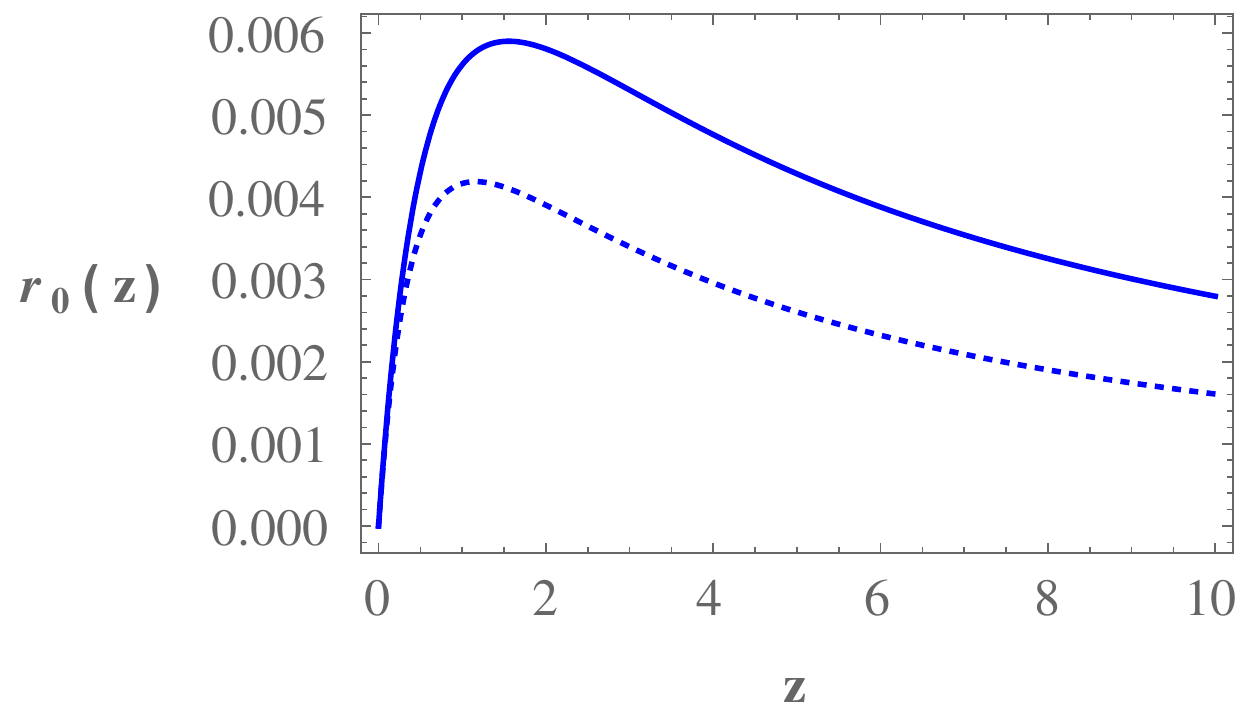}\hspace{2mm}
\caption{{\footnotesize The behavior of the
observer area distance $r_0(z)$ in the units of $H_0^{-1}$ for null vector
fields with FLRW background in GR in the presence and
absence of the cosmological constant (the solid and the dotted curves, respectively.)
We have assumed $D=4$, $H_0=67.6 km/s/Mpc$, $\eta(0)=0$,
 $\frac{d\eta(z)}{dz}\mid_{z=0}=0.1$ and $8\pi G=1$.
Moreover, \{$\Omega_{m0}=0.31$, $\Omega_{r0}=0.01$, $\Omega_{\Lambda}=0.68$\} and
 \{$\Omega_{m0}=0.8$, $\Omega_{r0}=0.2$, $\Omega_{\Lambda}=0$\} are
 associated with the solid and dotted curves, respectively. }}
\label{r-GR-DD-LR}
\end{figure}

For the particular case where $D=4$, we have $_2F_1\left(1,\frac{4-D}{2};\frac{3}{2};\frac{\Omega_{m0}+\Omega_{r0}}
{\Omega_{m0}}\right)=1=_2F_1\left(1,\frac{4-D}{2};\frac{3}{2};
\frac{\Omega_{m0}+\Omega_{r0}(z+1)}{\Omega_{m0}}\right)$,
therefore equation \eqref{mattig-GD-D-0} reduces to
\begin{eqnarray}
r_0(z)=\frac{2}{H_0 \Omega_{m0}(z+1)} \left\{\Omega_{m0}+\Omega_{r0}
-\left[\frac{ (\Omega_{m0}+\Omega_{r0})(\Omega_{m0}+\Omega_{r0}+\Omega_{r0} z)}{z+1}\right]^{\frac{1}{2}}\right\}.
\label{mattig-GD-4}
\end{eqnarray}
Moreover, concerning the case (ii), for the sake of comparison with
the physical $\Lambda$CDM model (just as an mathematical exercise),
by respecting the equality $\Omega_{m0}+\Omega_{r0}+\Omega_{\Lambda}=1$,
we have also plotted the behavior of the observer area distance, see figure \ref{r-GR-DD-LR}.

(iii) For another particular case where only $\Omega_{r0}$ vanishes,
the behavior of $\eta(z)$ and $r_0(z)$ are shown by the dotted curves
in left and right panels of figure \ref{eta-r-GR-DD}, respectively.

\section{GDE for null vector field in the context of the BD theory}
\label{D-dim BD}

In this section, we would like to apply the GDE (obtained in Section \ref{OT-solution})
for some well-known exact solutions
in the context of the BD theory. Let us first investigate the energy conditions associated
with our herein FLRW-BD model. Moreover, we will establish two different
dynamical systems associated with the BD
cosmological model presented in Section \ref{OT-solution}.
However, for the sake of continuity in the content of this paper, let us
present them in \ref{dynamical}.

\subsection{Energy conditions in BD theory of gravity}
\label{EC}
In Einstein equations, the energy momentum tensor, on the right hand side, determines
how the spacetime is curved and hence how the gravity acts. Accordingly, any conditions
imposed on the matter-energy distribution immediately affect on the corresponding
conditions associated with the geometry in the left hand side.
In this respect, the matter-energy distribution
establish the causal and geodesic structures of
spacetime \cite{HEbook73, Wald.book84,CLM14}.
Such reasons have encouraged researchers to investigate the energy
conditions not only in the standard cosmology but
also in models established based on the alternative theories.

In this paper, following \cite{SW13,MM20}, and considering equations \eqref{fun-Fri-1}-\eqref{fun-Fri-2}, we have
shown that the energy conditions associated with the BD cosmology in $D$-dimensions are given by
\begin{eqnarray}\label{NEC}
{\rm NEC:}\,\, \,\,  \rho_{\rm tot}\!\!&+&\!\! p_{\rm tot}\geq0,\\
\label{WEC}
{\rm WEC:} \,\,\,\,   \rho_{\rm tot}\!\!&+&\!\! p_{\rm tot}\geq0 ,\hspace{10mm} {\rm and}  \hspace{10mm}   \rho_{\rm tot}\geq 0,\\
\label{SEC}
{\rm SEC:} \,\,\,\,   \rho_{\rm tot}\!\!&+&\!\! p_{\rm tot}\geq0, \hspace{10mm}  {\rm and} \hspace{10mm}   (D-3)\rho_{\rm tot}+(D-1)p_{\rm tot}\geq0,\\
\label{DEC}
{\rm DEC:}\,\, \,\,   \rho_{\rm tot}\!\!&\pm&\!\! p_{\rm tot}\geq0, \hspace{10mm} {\rm and} \hspace{10mm}   \rho_{\rm tot}\geq 0.
\end{eqnarray}

Using equations \eqref{ro-pi-total}, \eqref{rho-phi-gen} and \eqref{p-phi-gen}, the energy conditions
in our models presented in Section \ref{OT-solution} are given by
\begin{eqnarray}\label{NEC1}
{\rm NEC:}\!\!\!\!\!\!&&\rho+p+\omega\left(\frac{\dot{\phi}^2}{\phi}\right)-H\dot{\phi}+\ddot{\phi}\geq0,\\
\label{WEC1}
{\rm WEC:} \!\!\!\!\!\!&&\rho +\frac{\omega}{2}\left(\frac{\dot{\phi}^2}{\phi}\right)
+\frac{V(\phi)}{2}-(D-1)H\dot{\phi}\geq 0,\hspace{24mm} {\rm and} \hspace{8mm} {\rm NEC} ,\\
\label{SEC1}
{\rm SEC:}\!\!\!\!\!\!&&(D-3)\rho-V(\phi)+(D-2)\omega\left(\frac{\dot{\phi}^2}{\phi}\right)
+(D-1)\left(p+H\dot{\phi}+\ddot{\phi}\right)\geq 0,\hspace{3mm} {\rm and} \hspace{3mm} {\rm NEC},\\
\label{DEC1}
{\rm DEC:}\!\!\!\!\!\! &&\rho-p
+V(\phi)-(2D-3)H\dot{\phi}-\ddot{\phi}\geq 0,\hspace{26mm} {\rm and} \hspace{8mm} {\rm WEC}.
\end{eqnarray}

\subsection{Modified Sen-Seshadri exact solutions in D dimensions}
\label{Sen-Seshadri}

In order to solve the equations \eqref{fun-Fri-1}-\eqref{BD-FRW4}
and \eqref{BD-FRW3}, following \cite{SS03}, let us assume the power law solutions as\footnote{It is important to note that,
due to a mistake sign in equation (7) of \cite{SS03}, most of our herein
equations/relations in the particular case where $D=4$ do not reduce to those
obtained in that paper. Therefore, before moving to obtain the GDE associated with
 herein model, we will present some details regarding the exact
 solutions as well as the allowed ranges of the parameters of this model.}
\begin{eqnarray}
\label{BD-a}
a(t)={a_0} \left(\frac{t}{t_0}\right)^{r },\\\nonumber\\
\label{BD-phi}
\phi(t)={\phi_0} \left(\frac{t}{t_0}\right)^{s },
\end{eqnarray}
where $a_0$ and $\phi_0$ are the values of the corresponding quantities
at $t=t_0$; $r$ and $s$ are two parameters. Solutions \eqref{BD-a} and \eqref{BD-phi}
lead to take the energy density, pressure and the scalar potential as
\begin{eqnarray}
\label{ans1}
\rho(t)=\rho_c t^{s -2}, \\
\label{ans2}
p(t)=p_c t^{s -2},\\
\label{ans3}
V(\phi)=V_c \phi^{\frac{s -2}{s}}.
\end{eqnarray}
Substituting $a(t)$, $\rho(t)$ and $p(t)$, respectively, from \eqref{BD-a}, \eqref{ans1}
and \eqref{ans2} into \eqref{BD-FRW4}, we can obtain a relation between $\rho_c$ and $p_c$.
Another relation between them can be also obtained via
employing a combination of the equations \eqref{fun-Fri-1}
and \eqref{fun-Fri-2} in which $V(\phi)$ should be removed. Therefore, $\rho(t)$
and $p(t)$ are obtained. Finally, employing the later relations together with \eqref{ans3}
in either \eqref{fun-Fri-1} or \eqref{fun-Fri-2}, we obtain $V_c$. In summary, we get
\begin{eqnarray}
\label{rho-c}
\rho_c\!\!\!&\equiv&\!\!\!\frac{(D-1)\phi_0 r    \left[ s ^2(\omega+1)-s(r +1)- r(D-2)\right]}{(s -2){t_0}^{s }}, \\\nonumber\\
\label{p-c}
p_c\!\!\!&\equiv&\!\!\!-\frac{\phi_0  \left[s +r  (D-1)-2\right] \left[ s ^2(\omega+1)-s(r +1)- r(D-2)\right]}{(s -2){t_0}^{s }},\\\nonumber\\
\label{V-c}
V_c\!\!\!&\equiv&\!\!\!\frac{s  \phi_0 ^{2/s } \Big\{ (D-1)(Dr -2)r-s  \omega  \left[s +2 r  (D-1)-2\right]\Big\}}{(s -2){t_0}^{2}}.
\end{eqnarray}
We should note that the quantities $\rho_c$ and $p_c$ are only
two symbols to write the equations \eqref{ans1} and \eqref{ans2}
in compact forms (they should have been replaced by other symbols). Concretely, they do not have the same
dimension of the energy density, pressure and the potential, but instead,
according to the units chosen in this paper (i.e., $8\pi G_0=1=c$),
we get $[\rho_c]=[p_c]=L^{2-s}([\rho]=[p]=[V])$.

Substituting relations \eqref{BD-a}-\eqref{V-c}
to \eqref{rho-phi-gen} and \eqref{p-phi-gen}, we easily obtain
\begin{eqnarray}
\label{SS-rho-phi}
\rho_\phi\!\!\!&=&\!\!\!\frac{(D-1) \phi_0 r  s }{2 (s -2)t_0^s}\Big[ D r  -2s ( \omega +1) +2\Big]t^{s -2},
 \\\nonumber\\
\label{SS-p-phi}
p_\phi\!\!\!\!\!\!&=&\!\!\!\!\!\!\frac{\phi_0 s  }{2 (s -2)t_0^s} \left\{2 r  \left[3-s  (\omega +2)\right]
-D(D-1)r ^2  +2\left(s+Dr-2\right)\left[s  (\omega +1) -1\right] \right\}t^{s -2},
\end{eqnarray}
which yields
\begin{eqnarray}
\label{SS-W-phi}
W_\phi \equiv \frac{p_\phi}{\rho_\phi}=\frac{1}{D-1}\left[\frac{(D-2)(s -2) }
{Dr  -2 (s  \omega +s -1)}-\frac{(D-1)r+(s   -2)}{r }\right].
\end{eqnarray}

Before proceeding our discussions concerning the GDE as well as determining
the allowed regions for the parameters of our model, let us present an
important note about the energy density, pressure as well the EoS
parameter associated with the BD scalar field.
As mentioned, by imposing the conservation law for the ordinary matter in the BD
theory as well as using the Bianchi identity, it is easy to show
that $\rho_{\phi}$ and $\rho_{\phi}$ do not satisfy the
conservation law, see equation \eqref{cons-phi-matt}.
Concretely, it sounds that $W_{\phi}$ cannot be considered as a physically meaningful quantity.
It has been suggested that a suitable candidate should come from an actual measurable
quantity. For the FLRW model, admitting that the $H$ is a measurable quantity, it is
straightforward to show that $\tilde{W}_\phi\equiv \tilde{p}_\phi/\tilde{\rho}_\phi$
where (for more details in four-dimensions, see \cite{Faraoni.book, T02})
\begin{eqnarray}\label{pho-tild-phi}
 \tilde{\rho}_\phi&=&\rho_\phi+\frac{1}{2} (D-1) (D-2)H^2 (1-\phi ),\\
  \tilde{p}_\phi&=&p_\phi-(D-2)\left[ \frac{(D-1)}{2} H^2 +\dot{H}\right](1-\phi ),
\end{eqnarray}
can be introduced as a real EoS parameter. More concretely, now
$ \tilde{\rho}_\phi$ and $\tilde{p}_\phi$ satisfy the conservation law:
\begin{eqnarray}
\label{cons-phi-tild-matt}
\dot{\tilde{\rho}}_\phi+(D-1) H (\tilde{p}_\phi+\tilde{\rho}_\phi)=0.
\end{eqnarray}

We should note that with the new energy density $\tilde{\rho}_\phi$,
the Friedmann equation \eqref{fun-Fri-1} is rewritten as
\begin{eqnarray}
\label{new-Fri}
\frac{(D-1)(D-2)}{2}H^2=\rho+\tilde{\rho}_\phi.
\end{eqnarray}
 Consequently, since the tilde quantities are directly related to the
 observable Hubble parameter and as they satisfy the conservation law, we
 therefore can consider them for comparing with the predictions of GR \cite{T02}.
 It should be noted that the discrepancy between $\tilde{W}_\phi$ and ${W}_\phi$ is very
 small \cite{T02}. Therefore, for simplicity, let us proceed
 our investigation with only the non-tilde quantities.

In order to obtain the solutions which could describe the present universe,
the range of parameters of the model should be restricted as follows.
(i) The range of the deceleration parameter $q=-a\ddot{a}/\dot{a}^2$ should be
restricted such that it would match with the recent observational data \cite{PL.CosPar18}.
For the power-law solution \eqref{BD-a}, we have $q=1/r-1$. Therefore, for an
 accelerating scale factor, $r$ must be greater than one.
(ii) From relations \eqref{rho-c} and \eqref{p-c}, the EoS parameter
associated with the ordinary matter (as a perfect fluid ), $W\equiv p/\rho$, is:
\begin{eqnarray}\label{DB-W}
W=\frac{2-s }{(D-1)r  }-1,
\end{eqnarray}
which should be constrained as $0<W<1$. Therefore, we obtain
\begin{eqnarray}\label{DB-s-range}
2 \left[1-(D-1)r\right]<s <2-(D-1)r.
\end{eqnarray}
From equation \eqref{DB-s-range}, for $r>1$ and $D\geq3$ we get $s<0$.

(iii) The energy densities should take positive values. Therefore, by taking
into account the conditions $r>1$ and $s<0$, from relations \eqref{rho-c} and
\eqref{SS-rho-phi}, we get the following allowed range for the BD coupling parameter:
\begin{eqnarray}\label{DB-omega-range-1}
\frac{Dr+2(1-s)}{2 s }<\omega <\frac{(D-2)r +(r+1)s -s ^2}{s ^2}.
\end{eqnarray}
On the other hand, $\omega$ should also be restricted as
\begin{eqnarray}\label{DB-omega-range-2}
\omega >-\left(\frac{D-1}{D-2}\right),
\end{eqnarray}
to get positive values for the energy density associated
with the scalar field in the Einstein frame.

By substituting $\phi$, $\rho$ and $\rho_\phi$ from
relations \eqref{BD-phi}, \eqref{ans1} and \eqref{SS-rho-phi} to \eqref{density.par.def}, we obtain
\begin{eqnarray}
\label{DP-BD}
\Omega &=&\frac{2}{D-2}\left[\frac{ (\omega +1)s^2 -s(r+1)-(D-2)r  }{ r  (s -2)}\right],\\\nonumber\\
\label{DP-phi-BD}
\Omega_\phi& =&-\frac{2(\omega +1)s ^2 -s  (Dr  +2)}{(D-2)r  (s -2) }.
\end{eqnarray}

Employing relations \eqref{SS-W-phi}, \eqref{DP-BD} and \eqref{DP-phi-BD},
$W_\phi$ is written as
\begin{eqnarray}
\label{SS-W-phi-1}
W_\phi= \frac{q+1}{D-1} \left[2+s \left(\frac{ \Omega }{\Omega \phi }\right)\right]-1.
\end{eqnarray}

\begin{figure}
\centering{}\includegraphics[width=2.6in]{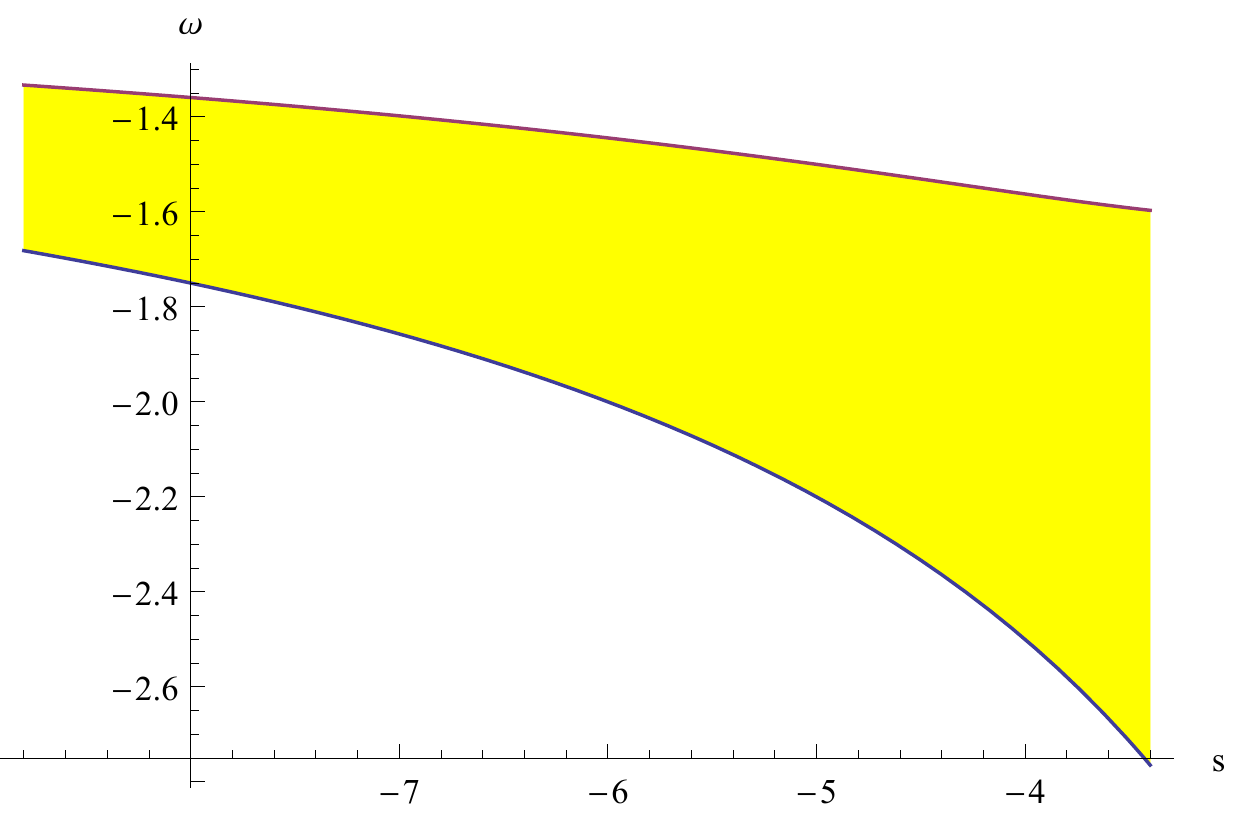}\hspace{2mm}
\caption{{\footnotesize  The allowed region of $(s,\omega)$ parameter
space for $r =2.5$. }}
\label{s-omega}
\end{figure}

 Using \eqref{DP-BD}, the allowed region of $(s,\omega)$ can be determined
 for some particular values of $r$ (or $q$) such that the range of $\Omega$ to be in accordance
 with observational data, see for instance figure \ref{s-omega}.

Let us check that whether or not the energy
conditions being satisfied for our herein BD cosmology.
Substituting the energy density and pressure from
relations \eqref{ans1}, \eqref{ans2}, \eqref{SS-rho-phi} and \eqref{SS-p-phi}
 into equations \eqref{NEC}-\eqref{DEC}, for arbitrary positive values of $t_0$ and $t$, we obtain
\begin{eqnarray}\label{NEC2}
{\rm NEC:}\,\, \,\, &&(D-2)\phi_0 r \geq0,\\
\label{WEC2}
{\rm WEC:} \,\,\,\, &&\frac{1}{2}(D-1)(D-2) \phi_0 r ^2\geq0,\hspace{18mm} {\rm and} \hspace{8mm} {\rm NEC} ,\\
\label{SEC2}
{\rm SEC:} \,\,\,\, &&-(D-1)(D-2)\phi_0r   \left(r-1\right) \geq0 ,\hspace{10mm} {\rm and} \hspace{8mm} {\rm NEC},\\
\label{DEC2}
{\rm DEC:}\,\, \,\,  &&(D-2)\phi_0r \left [(D-1)r-1\right]\geq0,\hspace{12mm} {\rm and} \hspace{8mm} {\rm WEC}.
\end{eqnarray}
It is seen that, assuming an attractive gravity (i.e., $\phi>0$)
and an accelerated scale factor, the
NEC, WEC and DEC are satisfied for our model.
However, in order to satisfy the SEC, inequality \eqref{SEC2}
yields $0\leq r\leq1$ which cannot be applicable for the present universe.

The GDE equation for this case is obtained by employing
relations \eqref{BD-a}-\eqref{SS-p-phi} in equation \eqref{ops5}:
\begin{eqnarray}
\frac{d^2\eta}{dz^2}+\frac{f}{(1+z)}\frac{d\eta}{dz}+\frac{f-2}{(1+z)^2}\eta=0,
\label{ops5-BD-2}
\end{eqnarray}
where $f\equiv2+1/r$ that can be written as
\begin{eqnarray}\label{f}
f&=& \frac{(D-1) (W_\phi +1)\Omega_\phi}{s  \Omega +2\Omega_\phi}+2.
\end{eqnarray}

An analytical exact solution for equation \eqref{ops5-BD-2} is
\begin{eqnarray}
\eta (z)= C_1 (z+1)^{\frac{(1-f-\left| f-3\right|)}{2}}
+C_2 \left(z+1\right)^{\frac{(1-f+\left| f-3\right|)}{2}},
\label{exact-BD-eta}
\end{eqnarray}
where the integration constants $C_1$ and $C_2$ carry the dimension of $\eta(z)$.

Moreover, by applying \eqref{mattig-1}, we obtain a general formula for the
observer area distance $r_0(z)$ associated with the power-law solution for our herein BD model:
\begin{eqnarray}\label{exact-BD-r}
r_0(z)=\frac{
\left[(z+1)^{\left| f-3\right| }-1\right](z+1)^{\frac{ (1-f-\left| f-3\right|)}{2}}}{H_0 \left| f-3\right| }.
\end{eqnarray}

In order to plot $\eta$ and $r_0$ in terms of $z$, we
should determine the allowed values of $f$ (which, in turn, is a function of $D$,
$\Omega$, $\Omega_\phi$ and $W_\phi$) and the integration constants $C_1$ and $C_1$.
In this regard, we apply the following
procedure: (i) we assume $\eta(0)=0$ and $d\eta(z)/dz\mid_{z=0}=0.1$.
Therefore \eqref{exact-BD-eta} yields
\begin{eqnarray}
C_1=-C_2=-\frac{0.1}{\left| f-3\right|}.
\label{IC-BD}
\end{eqnarray}

\begin{figure}
\centering{}\includegraphics[width=2.6in]{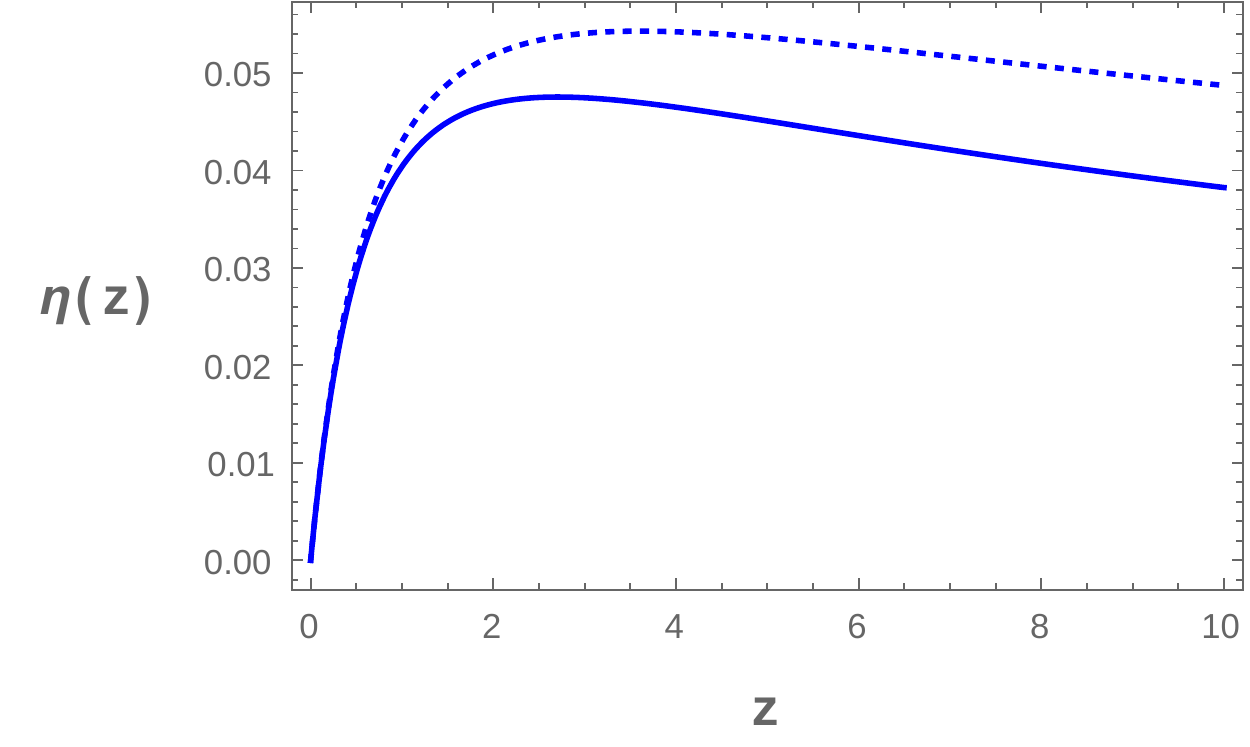}\hspace{2mm}
\includegraphics[width=2.6in]{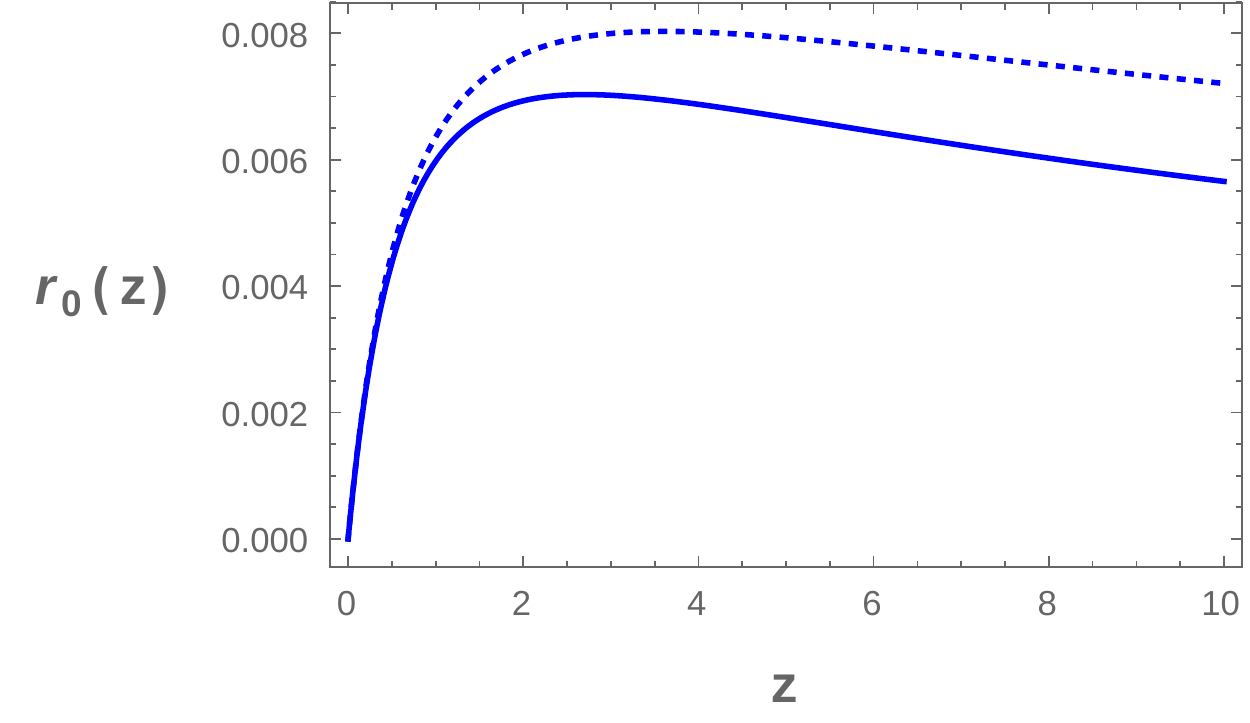}
\caption{{\footnotesize  The behavior
of the deviation vector $\eta(z)$ (the left panel) and the observer area
distance $r_0(z)$ (the right panel) in terms of the redshift parameter $z$ for null
vector fields with FLRW background in the context of the BD theory
(including a scalar field) in four dimensions. For all cases, we have
assumed $8\pi G=1$ and $\eta(0)=0$. In order to plot $\eta(z)$ and $r_0(z)$,
we assumed $d\eta(z)/dz\mid_{_{z=0}}=0.1$ and $ H_0=67.6 KM/s/Mps$, respectively.
Moreover, we have assumed $\Omega=0.32$, $\Omega_{\phi}=0.68$, $q\approx-0.43$
(the solid curves) and $q=-0.60$ (the dotted curves).}}
\label{eta-r-Sen-Seshadri}
\end{figure}

Consequently, with the obtained initial conditions and using the recent
observational data \cite{PL.CosPar18}, we could depict the behavior of the quantities of the model.
In figure \ref{eta-r-Sen-Seshadri}, we have plotted the behavior of the $\eta(z)$
and $r_0(z)$ for some allowed values of the corresponding parameters.

\subsection{Standard BD theory}
\label{standard-BD}

In this section, we restrict our attention to a cosmological application
in the context of the standard BD theory for which there is no scalar potential \cite{BD61}.
Moreover, let us assume that the cold matter with negligible pressure is
dominated. Therefore, letting $p=0$, equation \eqref{BD-FRW4} yields
\begin{eqnarray}
\rho a^{D-1}={\rm constant}.
\label{cons-ordi-2}
\end{eqnarray}


As our herein case is indeed a particular case of that presented in the previous
subsection, we can retrieve the corresponding solutions from those
obtained there. Therefore, using equation \eqref{p-c}, we obtain
\begin{eqnarray}
s=-(D-1)r  +2.
\label{BP-beta}
\end{eqnarray}
Moreover, $V_c$, given by equation \eqref{V-c}, should be set equal to
zero, we therefore get\footnote{Another solution is $r=\frac{2}{D-1}$ and
$s=0$. In this case, we obtain $\phi={\rm constant}$ and $a(t)\propto t^{2/(D-1)}$, which corresponds
to the matter dominated case in GR.}
\begin{equation}\label{BP-alfa}
r= \frac{2 (\omega +1)}{(D-1)\omega +D},
\end{equation}
where we have used \eqref{BP-beta}. Substituting $r$ and $s$ from
relations \eqref{BP-beta} and \eqref{BP-alfa} into \eqref{rho-c}, we get
\begin{equation}\label{BP-rho-c}
\rho_c=2 \left[\frac{(D-2) \omega +D-1}{(D-1) \omega +D}\right]\phi_0 t_0^{\frac{-2}{(D-1) \omega +D}}.
\end{equation}
Substituting $t=t_0$ to equation \eqref{ans1}  yields $\rho_c=\rho_0t_0^{2-s}$, hence
equation \eqref{BP-rho-c} can be rewritten as
\begin{equation}\label{BP-phi-0}
\phi_0=\frac{1}{2}\left[\frac{(D-1) \omega +D}{ (D-2) \omega +(D-1)}\right]\rho_0t_0^2.
\end{equation}
In summary, the solutions associated with this case are
\begin{eqnarray}
\label{BP-a}
a(t)&=&{a_0} \left(\frac{t}{t_0}\right)^{\frac{2 (\omega +1)}{(D-1) \omega +D } },\\\nonumber\\
\label{BP-phi}
\phi(t)&=&\frac{\rho_0t_0^2}{2}\left[\frac{(D-1) \omega +D}{ (D-2) \omega
+(D-1)}\right] \left(\frac{t}{t_0}\right)^{\frac{2}{(D-1) \omega +D} },\\\nonumber\\
\label{BP-rho}
\rho(t)&=&{\rho_0} \left(\frac{t}{t_0}\right)^{-\frac{2 (D-1) (\omega +1)}{(D-1) \omega +D} }.
\end{eqnarray}
In the particular case where $D=4$ these relations reduce to those obtained in \cite{BD61,BP01}.

Let us proceed our discussions considering the applicability of
this model for the late time for which we should have $q<0$.
Therefore, from \eqref{BP-a}, we obtain
\begin{eqnarray}
\label{BP-acc}
-\left(\frac{D-2}{D-3}\right)<\omega <-\left(\frac{D}{D-1}\right).
\end{eqnarray}

Before moving to focus on the GDE, we would discuss
concerning the energy conditions for this case.
Again, assuming an attractive gravity and arbitrary
values of the cosmic time, it is easy to show that
 these conditions are satisfied in the following ranges for the BD coupling parameter:
\begin{eqnarray}\label{3EC-BP}
{\rm NEC, WEC, DEC:}\,\, \,\, &\omega&  \leq -\left(\frac{D}{D-1}\right),
\hspace{12mm} {\rm or} \hspace{12mm}\omega\geq -1,\\\nonumber\\
\label{SEC-BP}
{\rm SEC:} \,\,\,\, &\omega&  \leq -\left(\frac{D-2}{D-3}\right) ,
\hspace{12mm} {\rm or} \hspace{12mm}\omega\geq -1.
\end{eqnarray}
Comparing inequalities \eqref{BP-acc}-\eqref{SEC-BP}, we find that if
we demand to get an accelerating scale factor, all the
energy conditions are satisfied except the SEC.

\begin{figure}
\centering{}\includegraphics[width=2.6in]{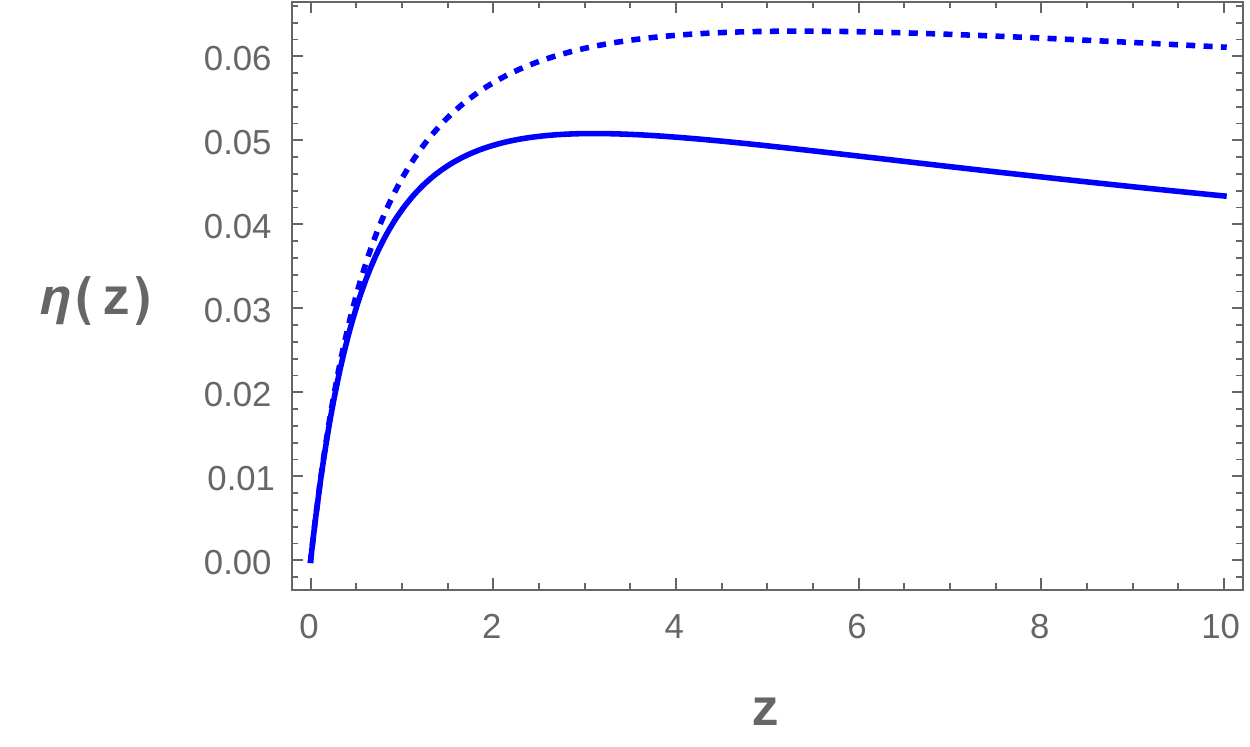}\hspace{2mm}
\includegraphics[width=2.6in]{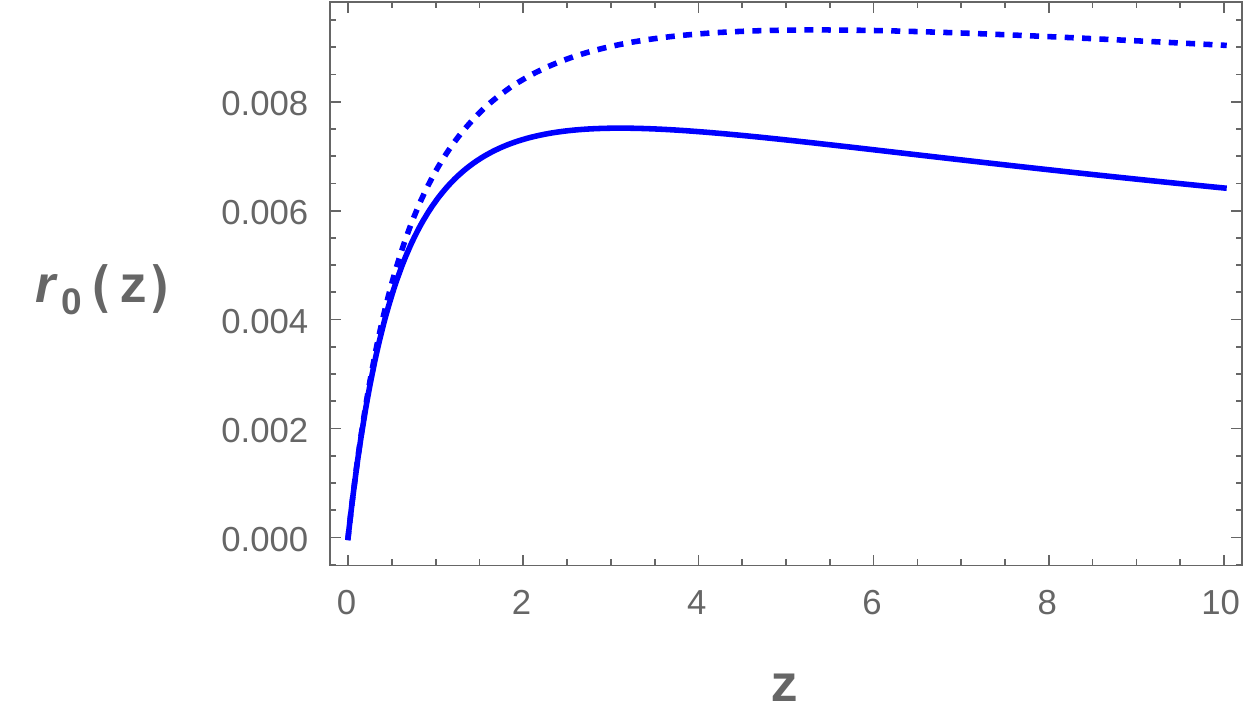}
\caption{{\footnotesize  The behavior
of the deviation vector $\eta(z)$ (the left panel) and the observer area
distance $r_0(z)$ (the right panel) in terms of the redshift $z$ for null
vector fields with FLRW background in the context of the standard BD theory
in four dimensions. For all cases, we have assumed $8\pi G=1$, $\eta(0)=0$.
In order to plot $\eta(z)$ and $r_0(z)$, we assumed $d\eta(z)/dz\mid_{_{z=0}}=0.1$
and $ H_0=67.6 KM/s/Mps$, respectively. The solid and dotted curves were plotted by assuming $\omega=-1.49$
 ($q\approx -0.52$) and $\omega=-1.40$ ($q=-0.75$), respectively.}}
\label{eta-r-BP}
\end{figure}

Concerning the GDE and its solution as well as the relation associated with the
 observer area distance, it is straightforward to show that all the
 relations \eqref{ops5-BD-2} and \eqref{exact-BD-eta}-\eqref{IC-BD} are
  still valid for this case. However, the relation of $f$ corresponding to this case is given by
\begin{eqnarray}
\label{BP-f}
f=\frac{(D+3) \omega +D+4}{2 (\omega +1)}.
\end{eqnarray}
For $D=4$, the constraint \eqref{BP-acc} reduces to $-2<\omega<-4/3$.
However, as mentioned, to get a positive energy density in the Einstein frame, the
BD coupling parameter should be restricted to $\omega>-3/2$.
Therefore, we should restrict our attention to a very narrow range $-3/2<\omega<-4/3$
(which yields a constraint on the deceleration parameter as $-1<q<-0.50$), for
which we get an accelerating scale factor that is applicable for the present
universe\rlap.\footnote{Although our herein power-law solutions yield accelerating
 universe, but they have their own problems, which will be discussed in Section \ref{Concl}. }
For $D=4$, by taking the allowed values of $\omega$ associated with the
accelerating late time universe, the behavior of $\eta(z)$ and $r_0(z)$ has
been plotted in figure \ref{eta-r-BP}.

\section{GDE for null vector field in the context of the MBDT}
\label{MBDT}

In what follows, let us present a brief description of the
MBDT in arbitrary dimensions (for more details, see \cite{RFM14}).\\


In analogy with \eqref{BD-Eq-DD} and \eqref{D2-phi}, we can write the BD field
equations in (D+1)-dimensional spacetime as
\begin{equation}\label{(D+1)-equation-1}
G^{^{(D+1)}}_{ab}=\frac{T^{^{(D+1)}}_{ab}}{\phi}+\frac{\omega}{\phi^{2}}
\left[(\overline{\nabla}_a\phi)(\overline{\nabla}_b\phi)-\frac{1}{2}{\cal G}_{ab}
(\overline{\nabla}^c\phi)(\overline{\nabla}_c\phi)\right]
+\frac{1}{\phi}\Big(\overline{\nabla}_a\overline{\nabla}_b\phi-{\cal G}_{ab}\overline{\nabla}^2\phi\Big),
\end{equation}
\begin{equation}\label{(D+1)-equation-4}
\overline{\nabla}^2\phi=\frac{ T^{^{(D+1)}}}{(D-1)\omega+D},
\end{equation}
 where we have not assumed any ad hoc scalar potential in $(D+1)$-dimensional spacetime (bulk).
 The Latin indices run from zero
to $D$; all the quantities with superscript $(D+1)$ and/or Latin
indices are associated with the bulk.
 Moreover, in order to recognize the covariant derivative in the bulk from
 its corresponding one on the hypersurface, we denote it with an overline, i.e., $\overline{\nabla}_a$.


 In Ref. \cite{RFM14}, by employing an appropriate reduction procedure, it
 has been shown that the effective field equations on
 a $D$-dimensional hypersurface can be obtained from
 \eqref{(D+1)-equation-1} and \eqref{(D+1)-equation-4}.
Concretely, it was assumed that the $(D+1)$-dimensional
spacetime is described by the metric
\begin{equation}\label{global-metric}
dS^{2}={\cal G}_{ab}(x^c)dx^{a}dx^{b}=
g_{\mu\nu}(x^\alpha,l)dx^{\mu}dx^{\nu}+
\epsilon\psi^2\left(x^\alpha,l\right)dl^{2},
\end{equation}

where $l$ is a
non-compact coordinate associated with the $(D + 1)$th
dimension; the scalar field $\psi$ depends on all coordinates;
$\epsilon=\pm1$ is an indicator allows the $(D + 1)$th
dimension to be either time-like or space-like.

By considering that the $(D+1)$-dimensional spacetime is
foliated by a family of $D$-dimensional
hypersurfaces $\Sigma$
(which, by taking $l=l_{0}={\rm constant}$, are denoted by $\Sigma_{0}$ ) being orthogonal to the
$(D+1)$-dimensional unit vector
\begin{equation}\label{unitvector}
n^a=\frac{\delta^a_{_D}}{\psi}, \qquad {\rm where} \qquad
n_an^a=\epsilon,
\end{equation}
along the extra dimension~\cite{Ponce1,Ponce2}, the induced metric $g_{\mu\nu}$ is
\begin{equation}\label{bulk-metric}
ds^{2}={\cal G}_{\mu\nu}(x^{\alpha},
l_{0})dx^{\mu}dx^{\nu}\equiv g_{\mu\nu}dx^{\mu}dx^{\nu}.
\end{equation}

In summary, it has been eventually shown that the MBDT is described by four sets of effective field
equations on the $D$-dimensional spacetime \cite{RFM14}:

\begin{enumerate}
  \item The equation associated with the scalar field $\psi$ is:

  \begin{eqnarray}\label{D2say}
\frac{\nabla^2\psi}{\psi}=\!\!\!\!\!&&-\frac{(\nabla_\alpha\psi)(\nabla^\alpha\phi)}{\psi\phi}
-\frac{\epsilon}{2\psi^2}
\left(g^{\lambda\beta}{\overset{**}g}_{\lambda\beta}+\frac{1}{2}{\overset{*}g^{\lambda\beta}}
{\overset{*}g}_{\lambda\beta}-\frac{g^{\lambda\beta}{\overset{*}g}_{\lambda\beta}{\overset{*}\psi}}{\psi}\right)
\\\nonumber
\!\!\!\!\!&&-\frac{\epsilon}{\psi^2\phi}
\left[\overset{**}\phi+\overset{*}\phi\left(\frac{\omega\overset{*}\phi}{\phi}
-\frac{\overset{*}\psi}{\psi}\right)\right]+\frac{1}{\phi}\left[\frac{(\omega+1)T^{^{(D+1)}}}{(D-1)\omega+D}
-\frac{\epsilon T^{^{(D+1)}}_{_{DD}}}{\psi^2}\right],
\end{eqnarray}
where $\overset{*}A\equiv \partial A/\partial l$.

 \item
 By introducing the quantity $P_{\alpha\beta}$ as
  \begin{equation}\label{P-mono}
P_{\alpha\beta}\equiv\frac{1}{2
\psi}\left(\overset{*}g_{\alpha\beta}
-g_{\alpha\beta}g^{\mu\nu}\overset{*}g_{\mu\nu}\right),
\end{equation}
it has been shown that in the MBDT, in contrary to the IMT, this
quantity is not conserved, but instead we obtain
\begin{equation}\label{P-Dynamic}
G^{^{(D+1)}}_{\alpha D}= \psi\nabla_\beta P^{\beta}{}_{\alpha}=\frac{T^{^{(D+1)}}_{\alpha D}}{ \phi}
+
\frac{\omega\overset{*}\phi(\nabla_\alpha\phi)}{\phi^2}
+\frac{\nabla_\alpha\overset{*}\phi}{\phi}
-\frac{\overset{*}g_{\alpha\lambda}\left(\nabla^\lambda\phi\right)}{2 \phi}-
\frac{\overset{*}\phi(\nabla_\alpha\psi)}{\psi\phi}.
\end{equation}

  \item
  Another couple of the field equations associated with the $D$-dimensional
   hypersurface, i.e., the counterpart equations corresponding
   to \eqref{(D+1)-equation-1} and \eqref{(D+1)-equation-4}, are
  given by \eqref{BD-Eq-DD} and \eqref{D2-phi}, respectively\footnote{It is important
  to note that equations \eqref{D2-phi} and \eqref{v-def} are the corrected
  versions of the corresponding ones presented in \cite{RFM14} (i.e.,
  equations (2.23) and (2.24)). Let us be more precise. In equation \eqref{D2-phi},
  the term $\phi \frac{dV(\phi)}{d\phi}$ is multiplied by the coefficient $(\frac{D-2}{d})$
  and therefore the wave equation can be obtained from the action \eqref{D-action}.
  We should note that for the particular case where $D=4$ (the main cosmological
  discussions of \cite{RFM14} have been presented in four dimensions) there are no
  discrepancies between these equations and those presented in \cite{RFM14}.
  In order to correct and to complete the results associated with $D\neq4$, let
  us investigate some additional topics in this paper.}.
  However, there is a main difference between the conventional BD theory
  (presented in Section \ref{OT-solution}) and the MBDT framework: in
  contrary to the former, the energy-momentum
  tensor $T_{\mu\nu}$ as well as the scalar potential $V(\phi)$ associated with the latter are not
  assumed by phenomenological assumptions, but instead they emerge
  from the geometry of the extra dimension.
Let us be more precise. (i) For the MBDT framework, the energy-momentum
tensor in equation \eqref{BD-Eq-DD} is given by $T_{\mu\nu}\equiv S_{\mu\nu}
+T^{^{\rm [MBDT]}}_{\mu\nu}$ where $S_{\mu\nu}$ and $T^{^{\rm [MBDT]}}_{\mu\nu}$ are given by
\begin{eqnarray}\label{S}
S_{\mu\nu}\equiv T_{\mu\nu}^{^{(D+1)}}-
g_{\mu\nu}\left[\frac{(\omega+1)T^{^{(D+1)}}}{(D-1)\omega+D}-
\frac{\epsilon\, T_{_{DD}}^{^{(D+1)}}}{\psi^2}\right],
\end{eqnarray}
\begin{eqnarray}\label{matt.def}
T^{^{\rm [MBDT]}}_{\mu\nu}\equiv T_{\mu\nu}^{^{[\rm IMT]}}+T_{\mu\nu}^{^{[\rm \phi]}}+\frac{1}{2}V(\phi)g_{\mu\nu}.
\end{eqnarray}
The three components of  $T^{^{\rm [MBDT]}}_{\mu\nu}$ in \eqref{matt.def} are:
\begin{eqnarray}\label{IMTmatt.def}
T_{\mu\nu}^{^{[\rm IMT]}}&\!\!\!\equiv &\!\!
\frac{\phi\nabla_\mu\nabla_\nu\psi}{\psi}
-\frac{\epsilon\phi}{2\psi^{2}}\left(\frac{{\overset{*}\psi}{\overset{*}g}_{\mu\nu}}{\psi}-{\overset{**}g}_{\mu\nu}
+g^{\lambda\alpha}\overset{*}{g}_{\mu\lambda}{\overset{*}g}_{\nu\alpha}
-\frac{1}{2}g^{\alpha\beta}\overset{*}{g}_{\alpha\beta}{\overset{*}g}_{\mu\nu}\right)\\\nonumber
\\\nonumber
&\!\!\!- &\!\!\frac{\epsilon\phi g_{\mu\nu}}{8\psi^2}
\left[{\overset{*}g}^{\alpha\beta}{\overset{*}g}_{\alpha\beta}
+\left(g^{\alpha\beta}{\overset{*}g}_{\alpha\beta}\right)^{2}\right],\\\nonumber
\\
T_{\mu\nu}^{^{[\rm
\phi]}}&\!\!\!\equiv &\!\!
\frac{\epsilon\overset{*}\phi}{2\psi^2}\left[\overset{*}g_{\mu\nu}
+g_{\mu\nu}\left(\frac{\omega\overset{*}\phi}{\phi}-g^{\alpha\beta}
{\overset{*}g}_{\alpha\beta}\right)\right],\label{T-phi}
\end{eqnarray}
and the term $V(\phi)$ is an induced scalar potential, which is obtained
from the following differential equation
\begin{eqnarray}\label{v-def}
\phi \frac{dV(\phi)}{d\phi}\!\!\!&\equiv&\!\!\! -2(\omega+1)
\left[\frac{(\nabla_\alpha\psi)(\nabla^\alpha\phi)}{\psi}
+\frac{\epsilon}{\psi^2}\left(\overset{**}\phi-
\frac{\overset{*}\psi\overset{*}\phi}{\psi}\right)\right]\\\nonumber
&&\!\!\!-\frac{\epsilon\omega\overset{*}\phi}{\psi^2}
\left[\frac{\overset{*}\phi}{\phi}+g^{\mu\nu}\overset{*}g_{\mu\nu}\right]
+\frac{\epsilon\phi}{4\psi^2}
\Big[\overset{*}g^{\alpha\beta}\overset{*}g_{\alpha\beta}
+(g^{\alpha\beta}\overset{*}g_{\alpha\beta})^2\Big]\\\nonumber
&&\!\!\!+2\left[\frac{(\omega+1)T^{^{(D+1)}}}{(D-1)
\omega+D}-\frac{\epsilon\, T_{_{DD}}^{^{(D+1)}}}{\psi^2}\right].
\end{eqnarray}

\end{enumerate}

In what follows, for later applications, let us mention a few comments.
\begin{itemize}

  \item
  Using the reduction procedure and the metric (\ref{global-metric}),
 the field equations (\ref{(D+1)-equation-1}) and (\ref{(D+1)-equation-4}) yield
four sets of effective equations (\ref{BD-Eq-DD}), (\ref{D2-phi}) (for which, the energy-momentum tensor and the
potential are obtained from the equations presented in this section rather
than phenomenological assumptions), (\ref{D2say})
and (\ref{P-Dynamic}) on the hypersurface.


\item

For a special case where $\omega\rightarrow\infty$ and/or
$\phi\sim 1/G_0$, the MBDT framework
 reduces to the IMT
 on a $D$-dimensional hypersurface, see \cite{RRT95} and references therein.
In order to obtain the equations obtained in \cite{RRT95} from
 those of the MBDT, in addition to the mentioned conditions, the induced
 scalar potential, without loss of generality, should be set equal to zero.
 Therefore, both sides of equation (\ref{D2-phi}) identically vanish.
 Consequently, equations \eqref{D2say}, \eqref{BD-Eq-DD} and \eqref{P-Dynamic} reduce to (see also \cite{RRT95})
 \begin{eqnarray}\label{D2say-IMT}
\epsilon\psi\nabla^2\psi=
-\frac{1}{2}g^{\lambda\beta}{\overset{**}g}_{\lambda\beta}-\frac{1}{4}{\overset{*}g^{\lambda\beta}}
{\overset{*}g}_{\lambda\beta}+\frac{g^{\lambda\beta}{\overset{*}g}_{\lambda\beta}{\overset{*}\psi}}{2\psi},
\end{eqnarray}
\begin{eqnarray}\label{IMT-Eq-DD}
G_{\mu\nu}^{^{(D)}}=T_{\mu\nu}^{^{[\rm IMT]}}, \hspace{15mm} \nabla_\beta P^{\beta}{}_{\alpha}=0.
\end{eqnarray}

\end{itemize}

In the rest of this section, not only we investigate the GDE in the context of the MBDT
framework, but also we present some modifications to \cite{RFM14} and
analyze new results. First, we should investigate
whether or not the GDE obtained in Section \ref{OT-solution} is also valid for the
MBDT framework. For this aim, we should only show that the induced matter on the
hypersurface whether or not obeys the equation of the perfect fluid.
Subsequently, as another application of the GDE, we will obtain exact solutions in the context of the MBDT.\\

Let us consider the metric of the ($D+1$)-dimensional spacetime as
\begin{equation}\label{ohanlon metric}
dS^{2}=ds^2+\epsilon \psi^2(t)dl^{2},
\end{equation}
where $ds^2$ is given by \eqref{brane-metric}.

For the cosmological problem in the context of the MBDT, we should restrict
our attention to the particular case: (i)  we assume there is no
ordinary matter in the bulk, i.e., $T^{^{(D+1)}}_{ab}=0$
(in order to justify such a specific choice, see, e.g., \cite{OW97},
and references therein); (ii) we impose the cylinder
condition \cite{OW97} (dropping the derivative of all quantities with respect
to the $(D+1)$-th coordinate $l$).
Consequently, from equations \eqref{D2say}, \eqref{matt.def}, \eqref{IMTmatt.def} and \eqref{v-def} we obtain
\begin{eqnarray}\label{cylinder-1}
T_{\mu\nu}\!\!&=&\!\!T^{^{\rm [MBDT]}}_{\mu\nu}= \frac{\phi\nabla_\mu\nabla_\nu\psi}{\psi}
+\frac{V(\phi)}{2}g_{\mu\nu}, \hspace{17mm}
T=\frac{\phi\nabla^2\psi}{\psi}+\frac{D}{2}V(\phi),\\\nonumber\\\nonumber
\\
\phi\,\frac{dV(\phi)}{d\phi}\!\!&\equiv&\!\! -2(\omega+1)
\frac{(\nabla_\alpha\psi)(\nabla^\alpha\phi)}{\psi},
 \hspace{24mm} \frac{\nabla^2\psi}{\psi}=-\frac{(\nabla_\alpha\psi)(\nabla^\alpha\phi)}{\psi\phi}.
\label{cylinder-2}
\end{eqnarray}

As the components of the induced energy-momentum tensor are essential to compute the GDE,
therefore, we let us obtain them. Using \eqref{ohanlon metric} and \eqref{cylinder-1},
we can easily show that the non-vanishing components of $T_{\mu\nu}$ are
\begin{eqnarray}\label{t-00}
\rho\equiv -T^{0}_{\,\,\,0}\!\!\!&=&\!\!\!
\phi\left[\frac{\ddot{\psi}}{\psi}-\frac{V(\phi)}{2\phi}\right],\\\nonumber
\\
\label{t-ii}
p_i\equiv T^{i}_{\,\,\,i}\!\!\!&=&\!\!\!-\phi\left[\frac{\dot{a}\dot{\psi}}{a\psi}-\frac{V(\phi)}{2\phi}\right].
\end{eqnarray}
where $i=1,2,...,(D-1)$, with no sum.

As it is observed, we have $p_i=p_j$ for any $i$ and $j$.
Namely, the induced energy-momentum tensor associated with the MBDT
can be assumed as a perfect fluid with energy density $\rho$ and pressure $p_i\equiv p$.
Consequently, the energy-momentum tensor
associated with our herein noncompact Kaluza-Klein framework can also be expressed by
relation \eqref{perfect}. Concretely, the formalism of the GDE presented throughout
Section \ref{OT-solution} (associated with the BD theory with a general scalar potential) are also valid
for the MBDT framework.
However, it is important to note that for the MBDT case \cite{RFM14},
in contrary to the phenomenological procedures used for the BD theory
as well as GR, the quantities $V({\phi})$, $\rho$ and $p$ have not
been chosen from employing some ad hoc assumptions, but instead we
have to use equations \eqref{cylinder-2}, \eqref{t-00} and \eqref{t-ii}.

Assuming $T^{^{(D+1)}}_{ab}=0$, equation \eqref{(D+1)-equation-4} yields a constant of motion as
\begin{equation}\label{cons-1}
a^{D-1}\dot{\phi}\psi=c_1,
\end{equation}
where $c_1\neq 0$ is an integration constant.
On the other hand, from solving equation \eqref{cylinder-2}, we get another constant of motion as
\begin{equation}\label{cons-2}
a^{D-1}\phi\dot{\psi}=c_2,
\end{equation}
where $c_2\neq0$ is a constant. Then, from combining equations \eqref{cons-1} and \eqref{cons-2}, we
easily obtain a relation between two scalar fields of the model as
\begin{equation}\label{cons-3}
\psi=C \phi^m, \hspace{10mm} m\equiv \frac{c_2}{c_1},
\end{equation}
where $C$ is an integration constant.
 In the context of the MBDT, we will see that obtaining a relation for $V(\phi)$ as
well as using relation \eqref{cons-3} assist us to rewrite the GDE equation in terms
of the BD scalar field and its derivatives with respect to the cosmic time.

Let us also relate the scalar field to the scale factor.
Equations \eqref{cylinder-1}, \eqref{cylinder-2} and \eqref{cons-3} yield
\begin{eqnarray}\label{T-time}
T&=&m\left(\frac{\dot{\phi}^2}{\phi}\right)+\frac{D}{2}V(\phi),\\
\frac{dV(\phi)}{d\phi}&=&2m(\omega+1)\left(\frac{\dot{\phi}}{\phi}\right)^2.
\label{v-phi}
\end{eqnarray}
On the other hand, for the metric \eqref{ohanlon metric}, we obtain
\begin{eqnarray}\label{D2-phi-2}
\nabla^2\phi=-\ddot{\phi}-(D-1)H\dot{\phi}.
\end{eqnarray}
Substituting quantities \eqref{T-time}--\eqref{D2-phi-2} into
equation \eqref{D2-phi}, and then using \eqref{cons-3}, we obtain
\begin{eqnarray}\label{D2-phi-3}
\ddot{\phi}+(D-1)H\dot{\phi}+\frac{m \dot{\phi}^2}{\phi}=0,
\end{eqnarray}
which can be written as
\begin{eqnarray}\label{D2-phi-4}
a^{D-1}\phi^m\dot{\phi}=c_3,
\end{eqnarray}
where $c_3$ is a constant.\footnote{Note that equation \eqref{D2-phi-4} can also be easily
obtained by substituting $\psi$ from \eqref{cons-3} to \eqref{cons-1}, and therefore $c_3=c_1/C$.
However, for the later applications of
equations \eqref{T-time}-\eqref{D2-phi-3}, we used the above approach.
Moreover, this approach may imply the consistency of the field equations of the MBDT.}


In order to obtain exact solutions for the field
equations, we first obtain the differential equation for the scale
 factor only: (i) We substitute
 $p_\phi$ and $p$ from relations \eqref{p-phi-gen} and \eqref{t-ii} into equation \eqref{fun-Fri-2}. Then,
 using \eqref{cons-3} and \eqref{D2-phi-4}, we obtain
\begin{eqnarray}
a^D\phi^{m+1}\Big\{ (D-2)a^D\phi^{m+1}\left[2a\ddot{a}+(D-3)\dot{a}^2\right]-2c_3(m+1)a^2\dot{a}\Big\}+c_3^2(\omega-2m)a^4=0.
\label{a-only1}
\end{eqnarray}
(ii) By adding equations \eqref{fun-Fri-1} and \eqref{fun-Fri-2}, and then
 employing a similar procedure used in (i), we obtain
\begin{eqnarray}\label{a-only2}
(D-2)\phi^{2(m+1)}(a\ddot{a}-\dot{a}^2)-c_3D(m+1)\phi^{m+1}a^{(2-D)}\dot{a}+c_3^2(\omega-2m)a^{2(2-D)}=0.
\end{eqnarray}
(iii) It is straightforward to show that combining
equations \eqref{a-only1} and \eqref{a-only2}, we can remove $\phi$:
 \begin{eqnarray}\label{a-only3}
\alpha (a\ddot{a})^2+\beta\dot{a}^2\left[2a\ddot{a}+(D-2)\dot{a}^2\right]=0,\hspace{3mm}
\alpha \equiv \omega-2m, \hspace{3mm}\beta\equiv D(m^2+\omega+1)-2\omega-(m-1)^2.
\end{eqnarray}
Equation \eqref{a-only3} can be rewritten as
\begin{eqnarray}\label{a-only4}
\alpha\left[\frac{d\left (\dot{a}\right)^2}{d\,{\rm lna}}\right]^2
+4\beta \dot{a}^2\left[\frac{d\left(\dot{a}\right)^2}{d\,{\rm lna}}+(D-2) \dot{a}^2\right]=0.
\end{eqnarray}

It is easy to show that there is a unique power-law solution
for equation \eqref{a-only3} as $a(t)=(R_1t+R_2)^r$, where $R_1$
and $R_2$ are integration constants and $r$ is a function of $\alpha$, $\beta$ and $D$.
Consequently, without loss of generality, we can take the
solutions of our herein cosmological model as
 \begin{eqnarray}
a(t)=a_0\left(\frac{t}{t_0}\right)^r,
\hspace{10mm} \phi(t)=\phi_0\left(\frac{t}{t_0}\right)^s,
\label{exact.sol-1}
\end{eqnarray}
where $r$ and $s$ are real parameters and $a_0$, $\phi_0$ are nonzero constants.
Therefore, $\psi(t)$ is obtained from \eqref{cons-3}:
 \begin{eqnarray}
\psi(t)=\psi_0\left(\frac{t}{t_0}\right)^n,\hspace{10mm} \psi_0\equiv CB^m t_0^n,\hspace{10mm} n=ms.
\label{exact.sol-psi}
\end{eqnarray}

We should note that the constants and parameters of
the model are not independent.
In order to obtain relations among them, substituting the scale factor
and the scalar field from \eqref{exact.sol-1} to equation \eqref{D2-phi-4}, we obtain
 \begin{eqnarray}
(D-1)r+s+n=1, \hspace{10mm} s\left(\frac{a_0}{t_0^r}\right)^{D-1}
\left(\frac{\phi_0}{t_0^s}\right)^{1+m}=c_3=\frac{c_1}{C}.
\label{exact.sol-m}
\end{eqnarray}
Moreover, using equation \eqref{fun-Fri-1}, we can
obtain a relation between the parameters $r$, $s$ and $\omega$ as
 \begin{eqnarray}
\omega=\frac{2 (n-1)(1-s)-D\left(n^2+s^2-1\right)}{(D-1) s^2}.
\label{exact.sol-omega}
\end{eqnarray}

Consequently, we can obtain exact expressions of the present
quantities of the model. For instance, substituting the scalar
field from \eqref{exact.sol-1} to equation \eqref{v-phi}, we
obtain the induced scalar potential in terms of $\phi$
as\footnote{There is also an induced scalar potential as a
logarithmic function, which, due to the problems
related to the energy conditions, will not be studied in this paper.}
\begin{eqnarray}\label{exact.sol-pot}
V(\phi)&=&\left[\frac{2\phi_0^{\frac{2}{s}}(\omega+1)
\,n\,s^2}{(s-2)t_0^2}\right]\phi^{\frac{s-2}{s}}.
\end{eqnarray}

Moreover, employing \eqref{exact.sol-1}, \eqref{exact.sol-m},
\eqref{exact.sol-omega} and \eqref{exact.sol-pot}, we easily
obtain the physical quantities in terms of the cosmic time:
\begin{eqnarray}\label{exact.sol-gen1}
\rho(t)\!\!&=&\!\!\frac{B n}{8 \pi  (s-2)}\Big\{ n (s-2)+2
-s\left[ (\omega +1)\,s +1\right]\Big\} t^{s-2},\\\nonumber\\\nonumber\\
\label{exact.sol-gen4}
V(t)\!\!&=&\!\!\frac{2B  (\omega +1)\, n\, s^2 \, t^{s-2}}{s-2},\\\nonumber\\\nonumber\\
\label{exact.sol-gen2}
W\!\!&=&\!\!\frac{n+1}{(D-1) r},
\\\nonumber\\\nonumber\\
\label{exact.sol-gen3}
\Omega\!\!&=&\!\!1-\Omega_\phi=\frac{2 (D-1) n \Big[n (s-2)-s (s \omega +s+1)+2\Big]}{(D-2) (s-2) (n+s-1)^2},
\\\nonumber\\\nonumber\\
\label{exact.sol-gen-3-1}
W_\phi \!\!&=&\!\!\frac{-n  \Big[2 Ds (\omega +2)-2 s (\omega +3)-4(D-2)\Big]+(s-2) \Big[(D-1)
 s\omega +2(s-1)\Big]}{(D-1) (2 n+s-2) \Big[s (\omega +2)-2\Big]},
\end{eqnarray}
where $\omega$ is given by \eqref{exact.sol-omega}.
Furthermore, using relations \eqref{exact.sol-omega} and
\eqref{exact.sol-gen2}, we obtain
\begin{eqnarray}
\label{r-MBDT}
r=\frac{2 \omega(W+1)  +3 W+2\pm \sqrt{  W^2+4 (\omega +1)
\left[W-(D-1)^{-1}\right]}}{ D \left[W^2 (\omega +2)+(2 W+1)
(\omega +1)\right]-(W+1) \left[W (\omega +2)+\omega \right]}.
\end{eqnarray}

Before moving to investigate the GDE for this model, let us
discuss concerning an important particular case where $\omega\longrightarrow \infty$.
Regardless of $W$, when the BD coupling parameter takes vary
large values, it is easy to show that \eqref{r-MBDT} reduces
to $r=\frac{2}{(D-1) (W+1)}$. Moreover, we obtain $s=0$
and $n\text=\frac{W-1}{W+1}$. Concretely, in this special
case, we retrieve $\phi={\rm constant}$, $V=0$ and the extra dimension
decreases with cosmic time for $-1<W<1$. Consequently, our
herein cosmological model for $|\omega|\longrightarrow \infty$ reduces
to the spatially flat FLRW model in GR in the absence
of the cosmological constant. As this case is not
applicable for the late time, we abstain from investigating the GDE for it.

Concerning the solutions for $\eta(z)$ and $r_0(z)$ associated with our herein model, it
is easy to show that all the equations \eqref{ops5-BD-2}-\eqref{IC-BD} are also valid for this case.
In what follows, let us retrieve the allowed ranges of the parameters associated with our herein cosmological model.

\begin{itemize}

  \item
   Letting $0<W<1$, we obtain $-1<n<(D-1)r-1$, which implies
that $n$ can take positive as well as negative values.
However, as we would like to have contracting extra
dimension \cite{OW97}, we therefore focus on the range $-1<n<0$.

  \item Imposing the positivity condition on $\rho$ and $\rho_\phi$, we obtain
\begin{eqnarray}\label{omega-MBDT}
\frac{2}{s}-2<\omega <\frac{n s-2 n-s^2-s+2}{s^2}, \hspace{3mm} -1<n<0,
\hspace{3mm} 2 \left[1-(D-1)r\right]<s <2-(D-1)r.
\end{eqnarray}

  \item
  Using equations \eqref{exact.sol-gen-3-1} and \eqref{omega-MBDT}, it is
straightforward to show that the allowed region of $W_\phi$ is:
\begin{eqnarray}\label{Wphi-MBDT}
\frac{(D-1) \iota s^2-2 (s-2) \left[D (s-1)-2 s+3\right]}{(D-1) \iota  (s-4) s}
<W_\phi <\frac{-D (s+2)+3 s+2}{(D-1) (s-2)},
\end{eqnarray}
where $0<\iota\ll 1$. Assuming $D=4$ and the allowed values
for $s$ (which is obtained from fixing the allowed ranges of $r$ and $n$), we
find that $W_\phi$ always takes negative values.
Therefore, letting $D=4$, taking a specific value
for $W_\phi$, and substituting the values of $\Omega$ from
observational data into equation \eqref{SS-W-phi-1}, we can
 easily determine the allowed range in the parameter space $(r,s)$.
 \end{itemize}

 The above procedure can also be used for obtaining the allowed
 values of $f$ in \eqref{f}, and we therefore can plot $\eta$ and
 $r_0$ in terms of the redshift parameter. Since, the behavior
 of these quantities is similar to those depicted
 in Section \ref{D-dim BD}, let us abstain from plotting them.

\begin{figure}
\centering{}\includegraphics[width=2.6in]{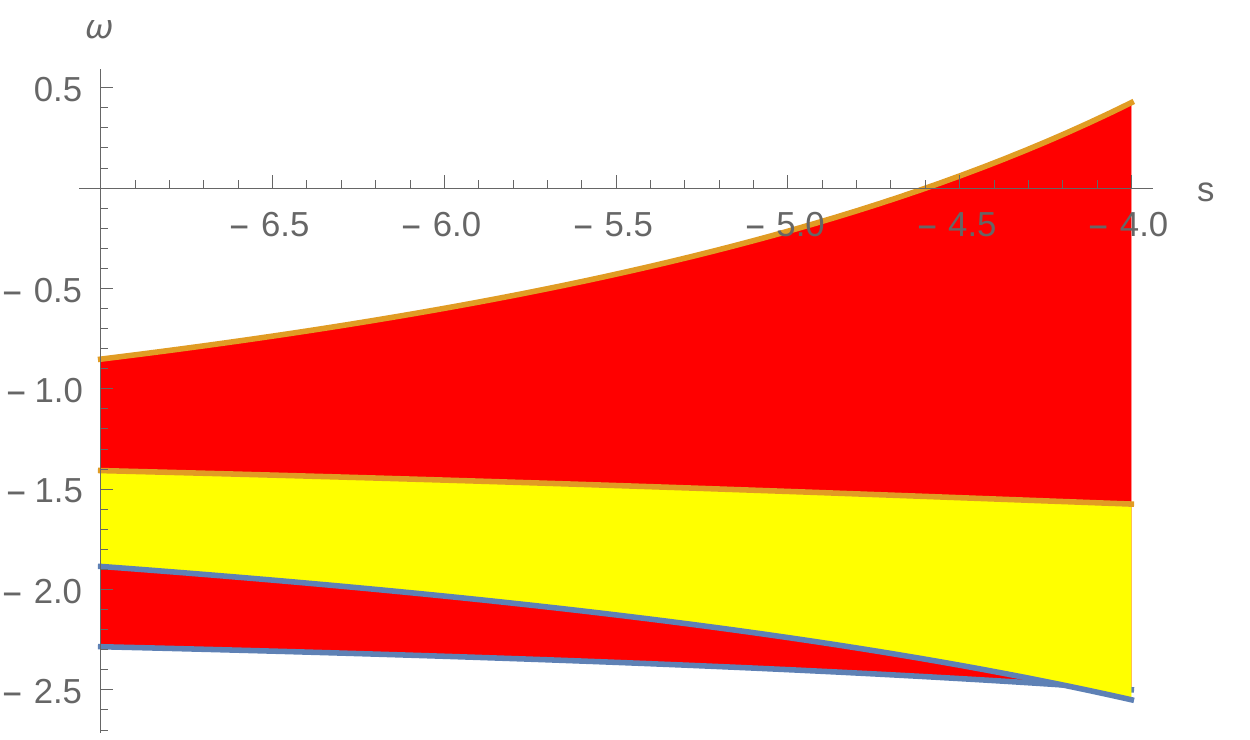}\hspace{2mm}
\caption{{\footnotesize  The allowed region of $(s,\omega)$ parameter
space for $r =2.6$ and $D=4$ associated with the MBDT model (whole the colored region)
 which includes that of the corresponding one for the conventional BD model (the yellow region only).}}
\label{omega-s-MBDT}
\end{figure}

Concerning the energy conditions, we should note that, for
our herein MBDT cosmological model, it is
straightforward to show that the conditions \eqref{NEC2}-\eqref{DEC2}
are also obtained for this case.
Namely, only the SEC is violated when the scale factor is accelerating.
In what follows, let us mention some advantages of our herein MBDT model with respect
to the BD model investigated in subsection \ref{Sen-Seshadri}.
 For the similar conditions, it is seen that the allowed
ranges for the BD coupling parameter in the MBDT is broader
than the corresponding one for the BD framework, see, for
instance, figure \ref{omega-s-MBDT}. In the MBDT cosmological
model, for every fixed value of $r$ (or $q$), we have another
 parameter, i.e., $n$ (the parameter associated with
  the extra dimension), which generates the range for $s$.
  For instance, for a particular case where $r=2.6$,
 $D=4$ and the allowed range of $n$, the allowed range of $s$ is
 specified as $-6.8<s<-5.8$, and therefore the allowed range
 of $\omega$ is shown in figure \ref{omega-s-MBDT}. Assuming that the
 BD coupling parameter must be greater than $-3/2$
 (for a four-dimensional spacetime), it is seen that
the allowed range of $\omega$, in the parameter space $(s,\omega)$, associated with
the conventional BD model is more narrower than the corresponding one of the MBDT model.

\subsection{Matter-dominated universe}
For this case, we can use the results obtained above. In order to
depict the behavior of $\eta(z)$ and $r_0(z)$, we should determine the allowed values
of $f$, or equivalently, of $s$ and $n$. For the matter dominated universe, we
should solve equation $W=0$. Therefore, from using \eqref{exact.sol-gen2} we obtain
\begin{eqnarray}
n=-1, \hspace{10mm}s=2+(1-D)r,
\label{s-mat}
\end{eqnarray}
where we used the relation \eqref{exact.sol-m}.
We should note that as $n<0$ therefore the extra dimension decreases with cosmic time.
 In order to compare our herein model with the corresponding case
 investigated in the context of the standard BD theory
 (see Subsection \ref{standard-BD}), let us express $r$ and $s$ in terms
 of the BD coupling parameter. Substituting $n=-1$ and $s$ from relation
 \eqref{s-mat} to \eqref{exact.sol-omega}, we can easily obtain such relations.
 In summary, the solutions associated with the matter-dominated case in the context of the MBDT are:
 \begin{eqnarray}\label{omega-r-mat}
 a(t)&=&a_0\left(\frac{t}{t_0}\right)^r,\hspace{10mm} r=\frac{2 \left[(D-1) (\omega +1)\mp \sqrt{(1-D) (\omega +1)}\right]}{(D-1) \left[(D-1) \omega +D\right]},\\\nonumber\\
 \label{omega-s-mat}
 \phi(t)&=&\phi_0\left(\frac{t}{t_0}\right)^s,\hspace{10mm}s=\frac{2 \left[1\pm\sqrt{(1-D) (\omega +1)}\right]}{(D-1) \omega +D},\\\nonumber\\
\label{omega-V-mat}
V&=&V_ct^{s-2},\hspace{15mm}V_c=\frac{4 B \left[(1-D) (\omega +1)\pm \sqrt{-(D-1) (\omega +1)}\right]}{(D-1) \left[(D-1) \omega +D\right]},\\\nonumber\\
\label{omega-rho-mat}
\rho &=&\rho_c t^{s-2},\hspace{15mm}\rho_c=\frac{B \left\{(D-1) \left[D (\omega +1)+1\right]\mp \sqrt{-(D-1) (\omega +1)}\right\}}{4 \pi  (D-1) \left[(D-1) \omega +D\right]},\\\nonumber\\
\label{omega-f-mat}
n&=&-1,\hspace{21mm} f=\frac{1}{2} \left[D+3\pm \frac{\sqrt{-(D-1) (\omega +1)}}{\omega +1}\right].
\end{eqnarray}
  Moreover, using \eqref{exact.sol-m}, \eqref{exact.sol-omega} and \eqref{s-mat}, the
  BD coupling parameter can be written as a function of $r$ and $D$. On the other
  hand, for the power-law solution \eqref{exact.sol-1}, $r$ is related to the
  deceleration parameter $q=-a\ddot{a}/(\dot{a})^2$ as
\begin{eqnarray}
r=\frac{1}{1+q}.
\label{r-q}
\end{eqnarray}
Consequently, we obtain $n$, $s$ and $\omega$ for the matter-dominated case as
 \begin{eqnarray}
n=-1, \hspace{7mm} s=2+\frac{1-D}{q+1},\hspace{7mm}
\omega =\frac{1-D}{(D-3-2 q)^2}-1.
\label{omega-q-mat}
\end{eqnarray}

Relations \eqref{omega-r-mat} and \eqref{omega-s-mat} imply that the BD
coupling parameter should be restricted as $\omega<-1$ and $\omega\neq-D/(D-1)$.
On the other hand, as mentioned, the BD coupling
parameter should be restricted
as $\omega >-(D-1)/(D-2)$ in $D$ dimensions.
Therefore, using this constraints on the relation of $\omega$ in \eqref{omega-q-mat}, we get
\begin{eqnarray}\label{q-allowed}
q>\frac{1}{2} \left(\sqrt{D^2-3 D+2}+D-3\right),\hspace{10mm}q<\frac{1}{2} \left(-\sqrt{D^2-3 D+2}+D-3\right).
\end{eqnarray}

We should note that the
inequalities \eqref{NEC2}-\eqref{DEC2} are also valid for
this case. Using
relations \eqref{omega-r-mat} and \eqref{r-q}, these constraints
can be written in terms of either $\omega$ or $q$.
In this case, we find that only the SEC is violated if we demand to get an accelerating universe.

\begin{figure}
\centering{}\includegraphics[width=2.6in]{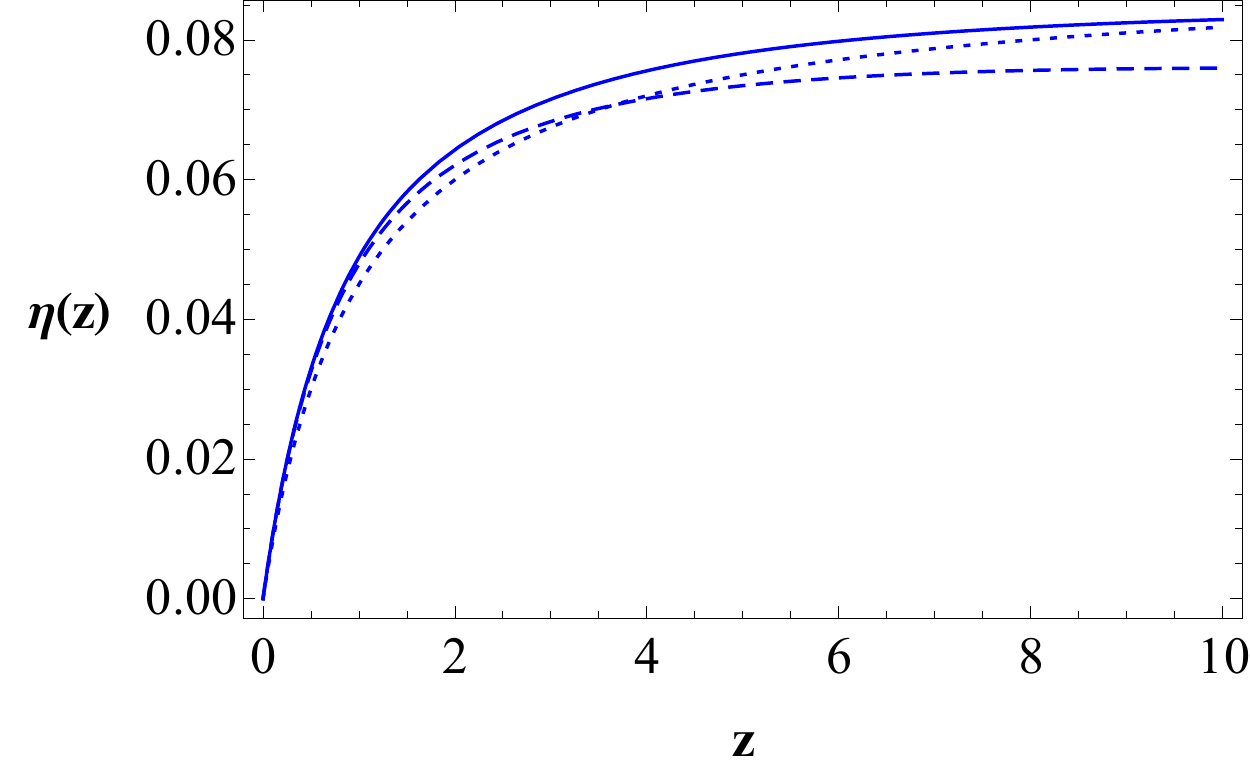}\hspace{2mm}
\includegraphics[width=2.6in]{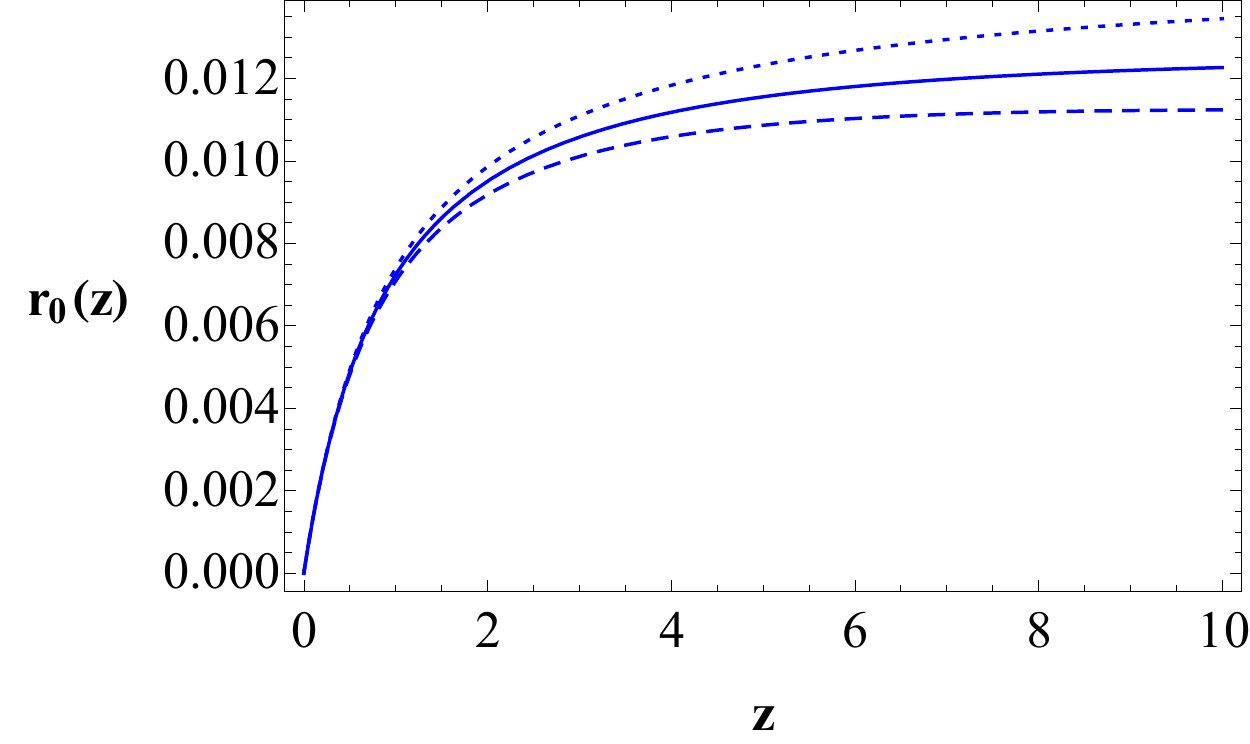}
\caption{{\footnotesize  The behavior
of the deviation vector $\eta(z)$ (the left panel) and the
observer area distance $r_0(z)$ (the right panel) in terms of the
redshift $z$ for null vector fields with FLRW background in four dimensions.
The solid, dashed and dotted curves are associated with the MBDT, standard BD
 and $\Lambda CDM$ models, respectively. For all cases, we have assumed $8\pi G=1$, $\eta(0)=0$ and
 a dust fluid with $W=0$. Therefore, for the $\Lambda CDM$
model, we have assumed $\Omega_{m0}=0=\Omega_{r0}$ and $\Omega_{\Lambda}=1$.
 In order to plot $\eta(z)$
and $r_0(z)$, we assumed $d\eta(z)/dz\mid_{_{z=0}}=0.1$ and
$H_0=67.6 KM/s/Mps$, respectively.
For both the standard BD and the MBDT, we have taken $\omega=-1.36$.}}
\label{eta-z}
\end{figure}

Let us focus on the four-dimensional universe. Substituting $D=4$, equation \eqref{q-allowed} yields
\begin{eqnarray}\label{q-allowed-2}
q>\frac{1}{2} \left(1+\sqrt{6}\right),\hspace{10mm}q<\frac{1}{2} \left(1-\sqrt{6}\right).
\end{eqnarray}
As the deceleration parameter for the present universe is
negative, therefore we restrict our attention to the
range $q<\frac{1}{2} \left(1-\sqrt{6}\right)$, which is in accordance with the
recent observational reports. Consequently, using relations \eqref{f},
\eqref{exact.sol-gen3}, \eqref{exact.sol-gen-3-1} and \eqref{omega-q-mat}, we finally obtain $f<(7-\sqrt{6})/2$.
For instance, in Fig. \ref{eta-z}, we have plotted the behavior
of the deviation vector $\eta(z)$ and the observer area distance
$r_0(z)$ in terms of the redshift $z$ (see solid curves).
As mentioned, our herein model in the limit $\omega\longrightarrow\infty$ is not
applicable for the present universe \cite{RFM14}. Therefore, for the sake
of comparison, we have also depicted the corresponding cases associated
with $\Lambda CDM$ and the standard BD model (discussed in Section \ref{GR}
and \ref{standard-BD}), see dotted and dashed curves, respectively.
It is important to note that our mathematical endeavors show that the behavior
of the depicted quantities for all the models are almost similar, see, for instance, figure \ref{eta-z}.

In what follows let us emphasize on anther great advantage of our
MBDT-FLRW model with respect to the corresponding one obtained in the BD framework.
As we have shown in this subsection (for the matter-dominated case), the
BD coupling parameter must be restricted as $-(D-1)/(D-2)<\omega<-1$.
However, for radiation, i.e., $W=1/(D-1)$, from \eqref{r-MBDT}, we obtain
\begin{eqnarray}
\label{r-MBDT-rad}
r=\frac{1+2 D (1+\omega)\pm 1}{D+D^2 (1+\omega)},
\end{eqnarray}
which implies that for the radiation-dominated case, $\omega$ is not restricted.
(For the upper sign, we obtain $r_+=2/D$ which is identical to
the corresponding case in GR.) This interesting result implies that the
reduced cosmology established in the context of the MBDT not
only yields an accelerating matter-dominated universe but also it gives
a consistent decelerating radiation-dominated universe. Let us compare
our herein MBDT model with that investigated in the context of the standard
BD theory (see Subsection \ref{standard-BD}): in the latter, the
accelerating matter-dominated epoch (where the restriction
on $\omega$ is given by \eqref{BP-acc}) is not consistent with the decelerating
radiation-dominated universe for which $\omega$ must be
restricted as $\omega>-D/(D-1)$ (in four dimension see \cite{BP01}.)
Therefore, in addition to the phenomenological approach employed in \cite{BP01},
another interesting way out of the inconsistency between the matter-dominated
and radiation-dominated epoches (faced in the standard BD theory) can be applying the MBDT framework.

\section{Conclusions and discussions}
\label{Concl}
In this paper, in analogy with four dimensional spacetime, the field
equations associated with the BD theory (including a general scalar potential)
were written in arbitrary dimension. Then we restricted our attention to
the spatially flat FLRW universe, which is filled with perfect fluid distribution,
and obtained the GDE corresponding to our model.
Subsequently, we focused on two particular cases, i.e., the fundamental
observers and past directed null vector fields, which yield
the Raychaudhuri equation and an extended
 versions of Mattig relation. Moreover, we have obtained a generalized
 expressions for the Ricci focusing condition, which implies that, in
 contrary to the corresponding GR case, it depends not only on the components of
 the ordinary matter fields but also on those associated
 with the BD scalar field, cf Section \ref{OT-solution}.

 For the particular case where the BD coupling parameter takes large
values, the BD scalar field and the potential being constants, we
have shown that the results of Section \ref{OT-solution} reduce to the
generalized version of those obtained for the $\Lambda CDM$ model.
Concerning the latter, we have studied the GDE for three cases, for which, by employing
either the exact analytical solutions or the numerical methods, we
have depicted the behavior of $\eta(z)$ and $r_0(z)$, cf Section \ref{GR}.

Assuming the spatially flat FLRW universe and the
 BD framework (in the presence of the scalar potential), we have obtained
generalized power-law exact solutions in $D$ dimensions, which are the
corrected as well as modified version of the well-known model investigated
in \cite{SS03}. This model for some different particular cases reduces to
those obtained in \cite{BD61,BM00}.
We should emphasize that one of our motivations in choosing the cosmological models with
a self-interaction potential $V(\phi)$ was to describe the epochs of the universe
 with small values of the redshift parameter, which, in turn, are suitable for
 applying our herein GDE for null vector field.

Moreover, we have investigated the energy conditions for our
herein BD cosmology and then applied them to the our exact solutions.
We have shown that, if we demand to apply these solutions for
the late time universe, only the SEC is violated.

It has been believed that
the cosmological models studied in Section \ref{D-dim BD}
face with a few problems: (i) The allowed range of the BD coupling
parameter that yields an accelerating scale factor associated with the
matter-dominated epoch cannot generate a consistent model
with the radiation-dominated duration.
An appropriate method to overcome to this problem can be
generalizing that model such that the constant $\omega$ is replaced by a varying
one, i.e., $\omega(\phi)$ \cite{BP01}.
(ii) Almost in all of the BD cosmological models, the acceleration
 as well as coincidence problems are solved provided that the $\omega$ takes
 low negative values, which is inconsistent with its lower
 limit imposed by solar system experiments \cite{BIT03}.
(iii) The components of the energy-momentum tensor associated with the BD
scalar field, in contrary to those of the ordinary matter, do not satisfy the
conservation law. In this respect, we defined other corresponding tilde
quantities, which not only are related to the measurable quantities of the
model but also satisfy the conservation law. However, due to very tiny
 discrepancy between the $W_\phi$ and ${\tilde{W}_\phi}$, we eventually proceeded
 our discussions with the un-tilde quantities.

For all the cosmological models presented in Section \ref{D-dim BD}, we
investigated the corresponding GDE for null vector field and depicted the
behavior of the quantities $\eta(z)$ and $r_0(z)$.

In order to have comprehensive discussions of the
GDE in the context of the BD models, we have also considered the modified
frameworks related to the modern Kaluza-Klein theories \cite{OW97}.
More concretely, the GDE has also been investigated in
the MBDT whose energy-momentum tensor and potential
are described in terms of geometrical quantities.
In this regard, we have shown that all the
results obtained in Section \ref{OT-solution} can also be applied for the MBDT framework.
 In order to analyze the GDE for null vector field for this case, we obtained the exact
 solutions where the components of the matter as well as the scalar
 potential emerge from the geometry of the extra dimension. We have
 mentioned the advantages of the MBDT-FLRW cosmology with respect to
 the corresponding one studied in the conventional BD theory.
 In this respect, we have also shown that the MBDT cosmological
  model, in contrast to the corresponding model
  obtained in the conventional BD framework, yields consistent
  radiation-dominated and matter-dominated epochs.

 Then,
 we have obtained the analytical exact solutions for $\eta(z)$ and $r_0(z)$, whose
 behavior have been depicted for this case and they have also been
 compared with those associated with GR and BD models.
 We have shown that the quantities $\eta(z)$ and $r_0(z)$ for all the mentioned models have
 similar behaviors for small values of $z$, as expected.

We have shown that for both the BD theory and the MBDT cosmological models
 in arbitrary dimensions (studied in \ref{Sen-Seshadri} and \ref{MBDT}),
 the Ricci forcing condition \eqref{BD-foc} yields $r>0$. Concretely, for
 an expanding universe in both of these models all families of null
geodesics experience focusing.
Moreover, assuming a vacuum fluid (cosmological constant) in the context of
the BD theory ($\rho+p=0$), the focusing condition leads to a
constraint on the parameters of the model as $s ^2 (\omega +1)-(r +1) s >0$.

In order to provide a complete description of the frameworks
presented in Sections \ref{OT-solution} and \ref{MBDT},
we have employed two different procedures to establish the most
general dynamical systems.
In the first approach, assuming a general power-law scalar potential,
we have obtained three coupled second order equations.
In some particular case, this dynamical setting reduces to those investigated
in \cite{KE95,HW98}. However, in the second approach, we did not impose any condition
 on the form of the scalar potential.
 Both of these frameworks can be appropriate
settings to analyze the BD-FLRW cosmological models.
Moreover, it worth noting that studying the evolution
of the density perturbation is important not only to check the stability of the model but also
 to compare the results of our herein cosmological models with their corresponding observational data.
 However, we should emphasize that investigating these structures have not been as the main
 objectives of the present paper. Therefore, we abstained from presenting a full
analysis of these interesting settings, but instead, in order to apply the
results of Section \ref{OT-solution}, we merely paid our attention to a
few particular classes of the exact solutions, which can be considered
as special subclasses generated by our herein dynamical systems.
Consequently, for the sake of continuity of the content
of this paper, we presented the general formalism of the dynamical systems in \ref{dynamical}.

\section{Acknowledgements}
We are very grateful to Prof. Anupam Mazumdar and K. Atazadeh
for their constructive comments. We would like to thank the anonymous
reviewers for their careful
reading of the paper and the valuable comments.
SMMR acknowledges the FCT grants UID-B-MAT/00212/2020 and UID-P-MAT/00212/2020 at CMA-UBI. Moreover, SMMR sincerely thanks Fatimah Shojai for giving the opportunity for
visiting the University of Tehran.
FS is grateful to the University of Tehran
for supporting this work under a grant provided by the
university research council.

\appendix
\section{Dynamical system approach for a generalized BD theory}
\label{dynamical}
The dynamical system theory has been employed as a powerful mathematical
tool for investigating the cosmological models within the context of
GR as well as alternative theories, see, for
instance \cite{WEbook97,RO89,F05,JK08,KJR14,SM16,PBBGP17,RB17,OO17,LZYLL19,GLPP20,SPL20} and
 references therein.
It has been believed that large classes of exact solutions
obtained in the context of BD theory could be recovered from a
corresponding suitable dynamical system, see for instance
\cite{RO89}. Obviously, any class of the exact solutions corresponding
to the set of equations \eqref{fun-Fri-1}-\eqref{BD-FRW4} are associated
with a special choice of the initial conditions. Therefore, presenting a
qualitative analysis of the dynamical system corresponding to a
cosmological model sounds to be a reasonable approach. Subsequently,
translating the results to the physically meaningful $a$-$\phi$ plane,
if possible, provides a complete description of the model.
In this respect, let us establish two dynamical settings for our herein cosmological mode.
However, investigating the dynamical scenario has not been in the scope of the
 present investigation, but instead, we have only restricted our attention to some well-known
 examples to observe whether or not our herein GDE does work well.
 In this regard, we present our dynamical systems in this section.

 In the first approach, using new variables, we transform the fourth-order
 system \eqref{fun-Fri-1}-\eqref{BD-FRW4} and \eqref{BD-FRW3} into three
 coupled second-order equations. Let use the
 conformal time as $d\tau=dt/a(t)$ and define the following variables
\begin{eqnarray}
\label{X}
X &\equiv & A\left(\frac{ \phi '}{\phi }\right),\hspace{25mm}{\rm where}\hspace{5mm}
A\equiv \frac{1}{D-2}\left[\frac{\chi(D)}{D-1}\right]^{\frac{1}{2}},\\
\label{Y}
Y&\equiv&\frac{a'}{a}+\frac{1}{D-2}\left(\frac{ \phi '}{\phi }\right),
\end{eqnarray}
where a prime denotes differentiation with respect to $\tau$.

Moreover, assuming a power-law scalar potential given by $V(\phi)=V_0\phi^\kappa$ and a barotropic
matter described by $p=W\rho$ (where $V_0$ is a
constant and $\kappa$, $W$ are two real parameters),
it is straightforward to show that
equations \eqref{BD-FRW3} and \eqref{fun-Fri-1}, respectively, transform to
\begin{eqnarray}
\label{dyn-phi-2}
X'\!\!\!&=&\!\!\!-(D-2) X Y -\frac{a^2 }{2 A (D-2)^2 (D-1)\phi}\Big\{ \left[(D-2)
\kappa-D\right]V_0 \phi ^{\kappa}+2 \rho  \left[(D-1)W-1\right]\Big\},\\\nonumber\\
\label{dyn-Fri}
Y^2\!\!\!&=&\!\!\!X^2+\frac{2 a^2 }{(D-1) (D-2) \phi }\left(\rho+\frac{V_0\phi^{\kappa}}{2}\right).
\end{eqnarray}

Finally, letting $Z\equiv V_0 a^2 \phi ^{\kappa-1}$, a generalized three-dimensional
dynamical system is retrieved for our herein model:
\begin{eqnarray}
\label{dyn-X}
X'\!\!\!&=&\!\!\!\left[\frac{(D-1) W-1 }{2 (D-2)A}\right]\left(X^2-Y^2\right)-(D-2) X Y
+\left[\frac{(D-1) (W+1)-(D-2) \kappa}{2 (D-1)(D-2)^2 A}\right]Z,\\\nonumber\\
\label{dyn-Y}
Y'\!\!\!&=&\!\!\!\frac{1}{2} (D-1) (W-1) X^2-\frac{1}{2}\left[(D-1)W+(D-3)\right]Y^2
+\frac{1}{2}\left(\frac{W+1}{ D-2}\right)Z,\\\nonumber\\
\label{dyn-Z}
Z' \!\!\!&=&\!\!\! \left\{\left[\frac{(D-2) (\kappa-1)-2}{  (D-2)A}\right]X+2Y\right\}Z.
\end{eqnarray}
For the spatially flat FLRW model, the
set \eqref{dyn-X}-\eqref{dyn-Z} can be considered as the most extended
case scenario obtained in the context of BD theory.
Let us mention a few special cases: (i) For $\kappa=2$ and $D=4$,
this model reduces to what investigated in \cite{SG97}. (ii)
Assuming $V_0=0$, the Z-equation disappears and we get a two-dimensional
dynamical system, which for $D=4$, we recover the model studied in \cite{KE95}.

Let us also provide the second dynamical system associated with our herein BD cosmological model.
Defining new variables as
\begin{eqnarray}
\label{dyn-x}
x\!\!\!&\equiv&\!\!\!\frac{1}{H}\left(\frac{\dot{\phi}}{\phi}\right),\\\nonumber\\
\label{dyn-y}
y\!\!\!&\equiv&\!\!\!\frac{1}{H}\sqrt{\frac{V\left(\phi\right)}{(D-1)(D-2)\phi}},\\\nonumber\\
\label{dyn-landa}
\lambda \!\!\!&\equiv&\!\!\! -\phi\left[\frac{dV\left(\phi\right)/d\phi}{V\left(\phi\right)}\right],
\end{eqnarray}
it is straightforward to show that the Friedmann equation and the acceleration equation can be written as
\begin{eqnarray}
\label{omega-acce}
\Omega\!\!\!&=&\!\!\!-\frac{\omega}{(D-1)(D-2)}x^2+\frac{2}{(D-2)}x-y^2+1,\\\nonumber\\
\label{dyn-y}
\frac{\dot{H}}{H^2}\!\!\!&=&\!\!\!-\frac{\omega}{(D-2)}x^2
-\frac{(D-1)[(D-2)(W+1)\omega+D]}{2\chi(D)}\Omega\cr\cr
&+&\left(\frac{D}{D-2}\right)x-\frac{(D-1)[(D-2)\lambda+D]}{2\chi(D)}y^2.
\end{eqnarray}
\newpage
Introducing $N\equiv lna$, we easily show that the following 3-dimensional dynamical
system can describe our herein D-dimensional BD model:
\begin{eqnarray}
\label{dyn-II}
\frac{dx}{dN}\!\!\!&=&\!\!\!-x\left(\frac{\dot{H}}{H^2}\right)-x^2-(D-1)x
+\frac{(D-1)(D-2)[1-(D-1)W]}{2\chi(D)}\Omega\cr\cr
&+&\frac{(D-1)(D-2)[(D-2)\lambda+D]}{2\chi(D)}y^2,\\\nonumber\\
\label{dyn-y}
\frac{dy}{dN}\!\!\!&=&\!\!\!-y\left[\frac{1}{2}(1+\lambda)x+\frac{\dot{H}}{H^2}\right],\\\nonumber\\
\label{dyn-landa}
\frac{d\lambda}{dN}\!\!\!&=&\!\!\!\left[(1-\Gamma)\lambda+1\right]x\lambda,
\end{eqnarray}
where
\begin{eqnarray}
\label{dyn-Gam}
\Gamma\equiv \frac{Vd^2V/d\phi^2}{\left(dV/d\phi\right)^2}.
\end{eqnarray}
In the particular case where $D=4$, the above dynamical system
reduces to that investigated in \cite{HS13} in which the dynamics for the specific case of the
quadratic scalar potential has been analyzed.

Moreover, if we assume that, in addition to the presence of
the scalar potential, the BD coupling parameter
depends also on the BD scalar field, the most generalized dynamical
system associated with the scalar-tensor theories is obtained.
A particular case of this model where $D=4$ and $V(\phi)=0$ has been established in \cite{RB17}.

Again, we emphasize that we abstained form
presenting complete analysis of the above dynamical systems in this paper.
Instead, we focused on a few specific exact solutions, which might be
generated from the above dynamical system formalisms.

\bibliographystyle{utphys}

\end{document}